\PassOptionsToPackage{table}{xcolor}
\documentclass{aastex701}

\newcommand{\finalevent}[1]{\textbf{#1}}
\begin{document}

\title{Helium-3 Enrichment in Gradual Solar Energetic Particle Events: Evidence for Jet-Supplied Seed Population}

\author[orcid=0000-0001-7381-6949, gname=Radoslav, sname=Bucik]{Radoslav Bu\v{c}\'{i}k}
\affiliation{Southwest Research Institute, San Antonio, TX 78238, USA}
\email[show]{radoslav.bucik@swri.org}  

\author[orcid=0000-0003-0508-4912,gname=Samuel, sname=Hart]{Samuel T. Hart} 
\affiliation{Southwest Research Institute, San Antonio, TX 78238, USA}
\email{samuel.hart@swri.org}

\author[orcid=0000-0001-9323-1200,gname=Maher, sname=Dayeh]{Maher A. Dayeh}
\affiliation{Southwest Research Institute, San Antonio, TX 78238, USA}
\affiliation{University of Texas at San Antonio, San Antonio, TX 78249, USA}
\email{maher.aldayeh@swri.org}

\author[orcid=0000-0002-7318-6008,sname=Mihir,gname=Desai]{Mihir I. Desai}
\affiliation{Southwest Research Institute, San Antonio, TX 78238, USA}
\affiliation{University of Texas at San Antonio, San Antonio, TX 78249, USA}
\email{mihir.desai@swri.org}

\author[orcid=0000-0003-2169-9618, gname=Glenn, sname=Mason]{Glenn M. Mason}
\affiliation{Applied Physics Laboratory, Johns Hopkins University, Laurel, MD 20723, USA}
\email{Glenn.Mason@jhuapl.edu}

\author[orcid=0000-0002-2825-3128, gname=Mark, sname=Wiedenbeck]{Mark E. Wiedenbeck}
\affiliation{Jet Propulsion Laboratory, California Institute of Technology, Pasadena, CA 91109, USA}
\email{mark.e.wiedenbeck@jpl.nasa.gov}

\setcounter{footnote}{0}

\begin{abstract}

Enhancements in ${}^3\mathrm{He}$ abundance, a characteristic feature of impulsive solar energetic particle (ISEP) events, are also frequently observed in gradual SEP (GSEP) events. 
Understanding the origin of this enrichment is crucial for identifying the mechanisms behind SEP generation. We investigate the origin of ${}^3\mathrm{He}$ enrichment in high-energy $(25\text{--}50\,\mathrm{MeV})$
solar proton events observed by SOHO, selecting events that coincide with $\lesssim 1\,\mathrm{MeV\,nucleon^{-1}}$ ${}^3\mathrm{He}$-rich periods 
detected by ACE during 1997--2021. The identified ${}^3\mathrm{He}$ enhancements include cases where material from independent impulsive (${}^3\mathrm{He}$-rich) 
SEP events is mixed with GSEP proton populations. Two high-energy proton events exhibit elemental composition and solar source characteristics consistent with ISEPs. Extreme-ultraviolet imaging
from SDO and STEREO reveals narrow, jet-like eruptions in the parent active regions of about 60\% of the remaining events. 
Notably, the highest ${}^3\mathrm{He}/{}^4\mathrm{He}$ ratios occur when coronal jets are present, consistent with fresh, jet-driven injection of suprathermal ${}^3\mathrm{He}$ that is subsequently 
re-accelerated during the event. Correspondingly, jet-associated events show fewer pre-event (residual) ${}^3\mathrm{He}$ counts, indicating that enrichment in these cases does not primarily 
come from remnant material. We find a positive correlation between ${}^3\mathrm{He}/{}^4\mathrm{He}$ and Fe/O, strongest in jet-associated events, consistent with a common jet-supplied seed population re-accelerated by the CME shock.

\end{abstract}

\keywords{\uat{Solar energetic particles}{1491} --- \uat{Solar abundances}{1474} --- \uat{Solar extreme ultraviolet emission}{1493} --- \uat{Solar active regions}{1974} }

\section{Introduction}   \label{sec:intro}

Suprathermal ions, traveling at speeds several times greater than the bulk solar wind, constitute the seed population for acceleration by coronal mass ejection (CME)-driven shocks in gradual solar energetic particle (GSEP) events \citep[e.g.,][]{2016LRSP...13....3D}. High-resolution abundance measurements from the Advanced Composition Explorer (ACE) spacecraft, launched in 1997, revealed that these events frequently exhibit modest enhancements in the abundances of ${}^3\mathrm{He}$ and $\mathrm{Fe}$, signatures associated with flare-related impulsive SEP (ISEP) events. ISEP events are characterized by extreme enhancements in the ${}^3\mathrm{He}/{}^4\mathrm{He}$ ratio, exceeding solar wind abundance by factors up to 10$^4$ \citep[e.g.,][]{2007SSRv..130..231M}. In addition, these events show significant enrichment in both heavy ions (Ne--Fe), by factors of 3--10, and ultraheavy ions $(\mathrm{mass} > 70\,\mathrm{AMU})$, by factors exceeding 100 in what appears to be a mass-per-charge dependence, though this dependence remains an active debate topic \citep{2021ApJ...908..243B,2023ApJ...957..112M,2024ApJ...974..220H,2025ApJ...981..178B}.

\citet{1999ApJ...525L.133M} examined 12 GSEP events and found that, in eight cases, the 0.5--2.0\,MeV\,nucleon$^{-1}$ ${}^3\mathrm{He}/{}^4\mathrm{He}$ ratio ranged from 5 to 135 times the average slow solar wind (SSW) value of $(4.08\pm0.25)\times10^{-4}$ \citep{1998SSRv...84..275G}. \citet{2006ApJ...649..470D} analyzed 64 GSEP events observed by NOAA GOES between November 1997 and January 2005, identifying 29 with distinct and finite ${}^3\mathrm{He}$ peaks. In these events, the 0.5--2.0\,MeV\,nucleon$^{-1}$ ${}^3\mathrm{He}/{}^4\mathrm{He}$ ratio was enhanced by factors of $\sim$2--150 over the SSW value, with only 23 (36\%) events exceeding a factor of 4. In that work, they cautioned that enhancements by a factor of $\sim$2 may reflect fluctuations in the solar wind ${}^3\mathrm{He}$ abundance; thus, only events with enhancements a factor $\geq$ 4 were considered statistically significant. A subsequent study by \citet{2016ApJ...816...68D}, covering 46 GSEP events from April 1998 to February 2014 (including 26 from the earlier survey), reported enhancements by factors $\sim$1.5--194 in 27 events, with 24 events (52\%) showing enhancements above a factor of 4. More recently, \citet{2023A&A...669A..13B} reported a GSEP event on 2020 November 24, observed by Solar Orbiter at 0.9\,au, with a ${}^3\mathrm{He}/{}^4\mathrm{He}$ enhancement of 24, following a prolonged period of ${}^3\mathrm{He}$-rich SEPs. Similar enhancements at higher energies ($>$5\,MeV\,nucleon$^{-1}$) have also been observed \citep[e.g.,][]{1999GeoRL..26.2697C,2000AIPC..528..107W}. 

Fe/O enhancements are frequently reported in these events \citep{1999ApJ...525L.133M,2006ApJ...649..470D,2016ApJ...816...68D,1999GeoRL..26.2697C,2005ApJ...625..474T}. While these may arise from magnetic rigidity-dependent transport effects \citep[e.g.,][]{2012ApJ...761..104M}, due to the lower charge-to-mass ratio of Fe compared to O, such transport biases do not account for ${}^3\mathrm{He}/{}^4\mathrm{He}$ enhancements \citep[e.g.,][]{1999ApJ...525L.133M}. Thus, ${}^3\mathrm{He}$ enrichment is a robust indicator of flare material in GSEP events \cite[e.g.,][]{2016ApJ...816...68D}.

Despite significant progress, the mechanisms by which CME-driven shocks access flare-accelerated suprathermal populations and generate the observed compositional signatures in GSEP events remain poorly understood. One proposed scenario involves the re-acceleration of remnant flare suprathermal ions, which are commonly present in the interplanetary (IP) medium at 1\,au \citep{1999ApJ...525L.133M,2005ApJ...625..474T,2006ApJ...649..470D,2012SSRv..171...97M,2014ASPC..484..234W,2023A&A...669A..13B}. An alternative scenario suggests the re-acceleration of suprathermal ions produced by flares in the parent active regions (ARs) \citep{2000AIPC..528..111V,2003GeoRL..30.8017C,2006JGRA..111.6S90C,2013ApJ...776...92K}. In addition to observational analysis, coupled flare and CME-driven shock acceleration has also been modeled \citep[e.g.,][]{2005GeoRL..32.2101L}. The presence of type III radio bursts in GSEP events indicates that electron beams escape along open field lines into IP space, suggesting that suprathermal ions from the same site can also escape \citep{2006JGRA..111.6S90C,2010JGRA..115.8101C}. In this context, the observer’s heliolongitude relative to the source flare has been suggested to influence the Fe/O ratio in GSEP events \citep{2003GeoRL..30.8017C,2013ApJ...776...92K}. 

In this study, we investigate the origin of ${}^3\mathrm{He}$ enhancements in ${}^3\mathrm{He}$-enriched GSEP events. We examine ion intensity time profiles for evidence of independent ISEP events. To evaluate the contributions of remnant suprathermal ions and suprathermals associated with concomitant activity, we analyze integrated ${}^3\mathrm{He}$ counts prior to the GSEP onset, magnetic connectivity, and the morphology of flare activity at the associated solar source regions.   

\section{Data and Event Selection}   \label{sec:data}

\subsection{Initial catalog}  \label{subsec:catalog}

We surveyed the live catalog of ${}^3\mathrm{He}$-rich periods detected by the Ultra-Low Energy Isotope Spectrometer \citep[ULEIS;][]{1998SSRv...86..409M} on ACE and compiled by \citet{2022ApJS..263...22H}, covering the interval from 1997 September 29 to 2021 March 1, with the aim of identifying concurrent GSEP events. The catalog lists periods in the energy ranges 0.32--0.45\,MeV\,nucleon$^{-1}$ and 0.64--1.28\,MeV\,nucleon$^{-1}$. For our analysis, we focused on the Solar Dynamics Observatory (SDO) era (post-May 2010), during which 174  ${}^3\mathrm{He}$-rich periods were reported. We selected proton events observed by Energetic and Relativistic Nuclei and Electron \citep[ERNE;][]{1995SoPh..162..505T} on the Solar and Heliospheric Observatory (SOHO) that exhibited sufficiently high intensities at higher energies. Specifically, we required that the ratio of the 25--50\,MeV proton intensity peak to the 24-hour pre-event background exceeded a factor of two. Such events exhibit visually distinguishable enhancements above background. In addition to GSEP events that overlap a cataloged ${}^3\mathrm{He}$-rich period, we also included six events for which a cataloged ${}^3\mathrm{He}$-rich period ended within 24 hours before the proton-event onset \citep[\#566, 604, 641, 682, 700, and 790;][]{2022ApJS..263...22H}. We treat this one-day window as an observational proxy for seed availability along the observer-connected field line. Notably, no ${}^3\mathrm{He}$ periods ended between one and two days before the onset, and periods ending three or more days earlier were excluded. 

A total of 41 proton events satisfied our selection criteria. Among these, 10 ${}^3\mathrm{He}$ periods (\#471, 537, 576, 646, 680, 681, 688, 744, 772, 794) showed no low-energy ($\lesssim 1\,\mathrm{MeV\,nucleon^{-1}}$) enhancements in ${}^4\mathrm{He}$ (and often O) above the preceding background and were therefore excluded. Period \#688 spanned two closely spaced proton events. We further excluded six ${}^3\mathrm{He}$ periods (\#560, 608, 672, 758, 773, 840) associated with corotating interaction regions (CIRs), identified by characteristic signatures in the solar wind plasma and IP magnetic field \citep[e.g., enhanced total pressure during gradual solar wind speed increases;][]{1995ISAA....3.....B}. One additional period (\#532) was excluded due to an observed bump associated with a forward IP shock, as noted in the shock database by \citet{2023FrASS..1040323O}. After all selections, 23 events remained for further analysis. 

\begingroup
\setlength{\tabcolsep}{3pt}     \renewcommand{\arraystretch}{1} 

\begin{deluxetable*}{cclllcclccccccl}
\tabletypesize{\scriptsize}
\tablecaption{Solar source characteristics of the proton events\label{tab:source}}
\tablewidth{0pt}
\tablehead{
\colhead{\#} & \colhead{\#} & \colhead{Proton event} & \colhead{Type II} & \colhead{Type III} &
\multicolumn{3}{c}{Flare} & \colhead{AR} & \colhead{$\Delta$} &
\colhead{Open to} & \multicolumn{2}{c}{CME} & \colhead{Jet} & \colhead{Ref.} \\ 
\cline{6-8}
\cline{12-13}
\colhead{} & \colhead{} & \colhead{start day} & \colhead{} & \colhead{} &
\colhead{Start} & \colhead{Class} & \colhead{Loc.} &
\colhead{} & \colhead{} & \colhead{ecl.} &
\colhead{Speed} & \colhead{Width} & \colhead{} & \colhead{} \\
\colhead{} & \colhead{} & \colhead{} & \colhead{(UT)} & \colhead{(UT)} &
\colhead{(UT)} & \colhead{} & \colhead{} &
\colhead{} & \colhead{(deg)} & \colhead{} &
\colhead{(km s$^{-1}$)} & \colhead{(deg)} & \colhead{} & \colhead{}
}

\decimals
\startdata
1 & 470 & 2010-Jun-12$^{\star}$ & 00{:}57{:}00 & 00{:}54{:}00$^{\mathrm{a}}$ & 00{:}30{:}00 & M2.0 & N23W48 & 1081 & 20  & Y$^{\dagger}$ & 486 & 119 & Y & i, j, k, l \\
2 & 473 & 2010-Aug-31$^{\ddagger,\star}$ & 21{:}00{:}00$^{\mathrm{b}}$ & 20{:}50{:}00 & \dots     & \dots & S22W147 & \dots & -80 & N$^{\mathrm{d}}$ & 1304 & 360 & Y & i, k, l \\
3 & 494 & 2011-Jan-28$^{\star}$ & 01{:}01{:}00 & 00{:}57{:}00 & 00{:}44{:}00 & M1.3 & N17W88 & 1149 & -17 & Y & 606 & 119 & N & i \\
4 & 508 & 2011-Apr-21$^{\ddagger,\star}$ & 00{:}38{:}00$^{\mathrm{b}}$ & 00{:}28{:}00$^{\mathrm{b}}$ & \dots & \dots & N18W144 & \dots & -98 & Y & 475 & 111 & Y & i \\
5 & 524 & 2011-Aug-02$^{\star}$ & 06{:}08{:}00 & 06{:}09{:}00 & 05{:}19{:}00 & M1.4 & N17W13 & 1261 & 38 & N$^{\mathrm{d}}$ & 712 & 268 & N & i \\
6 & 525 &	2011-Aug-08 &	18:03:00 & 18:03:00$^{\mathrm{a}}$ &	18:00:00 &	M3.5 &	N16W63	&1263 &	-24 &	N &	1343 &	237 &	N&i\\
7 & 539 &	2011-Nov-26$^{\star}$ &	07:09:00$^{\mathrm{a}}$  &	07:03:00 &	06:09:00 & C1.2 &	N11W48	&1353 &	8	&Y	& 933 &	360 &	N	& i, m\\
8 & 554 &	2012-Jan-19$^{\star}$  &	15:00:00$^{\mathrm{a}}$	& 14:39:00$^{\mathrm{a}}$  &	13:44:00  & M3.2	  &N32E27  &	1402  &	86	 &N$^{\mathrm{d}}$	  &1120 &	360 &	N &i\\
9 & 566 &	2012-May-17$^{\star}$ & 01:31:00 &01:33:00 & 01:25:00&	M5.1&	N07W77&	1476&	-14&	Y&	1582&	360&	N&	i, m\\
10 &571 & 2012-Jun-14$^{\star}$	&14:05:00$^{\mathrm{a}}$	&13:50:00$^{\mathrm{a}}$	&12:52:00&	M1.9	&S18E07	&1504&	73&	Y&	987&	360&	Y&	i\\
11 &	604 & 2013-Feb-26$^{\ddagger,\star}$&	10:20:00$^{\mathrm{b}}$&	10:05:00$^{\mathrm{b}}$&	\dots&	\dots	&N12W124&	\dots&	-52&	Y&	987&	360&	N&	i\\
12&	611& 2013-May-02&	05:06:00&		04:59:00&		04:58:00&		M1.1&		N11W26&		1731	&	27&		Y$^{\dagger}$ &		671&		99&		Y&		i, j, n\\
13&	641&	2013-Sep-30$^{\star}$ &21:53:00$^{\mathrm{a}}$&	21:55:00$^{\mathrm{a}}$	&21:43:00$^{\S}$ &	C1.2&	N10W33	&Filament	&50&	\dots&	1179	&360	&N	&i, j\\
14&	647&	2013-Oct-25$^{\star}$&07:59:00&	07:58:00 & 07:53:00 &	X1.7 &	S08E73 &	1882 	& 145& N$^{\dagger\dagger}$  &	587 &	360 &	Y & i\\
    &        &                                   &14:58:00&	14:58:00$^{\mathrm{a}}$	&14:51:00	& X2.1&	S08E70 &	1882	 & 142&	N$^{\dagger\dagger}$	&1081&	360 &	Y & i\\	
15&	665&	2013-Dec-26$^{\ddagger,\star}$	&03:02:00$^{\mathrm{c}}$ &	03:03:00$^{\mathrm{c}}$	& \dots	& \dots	& S09E162 &	\dots &	-119 &	Y &	1011 &	$>$171 &	N &	i\\
16&	668&	2014-Jan-04 &	18:55:00$^{\mathrm{a}}$	& 18:55:00$^{\mathrm{a}}$ &	18:47:00 &	M4.0 &	S11E28 &	1944 &	75 &	Y &	977 &	360 &	Y &	m\\
17&	682&	2014-Feb-25$^{\star}$ &	00:56:00 &	00:45:00 &	00:39:00 &	X4.9 &	S12E80 &	1990 &	137&	Y&	2147&	360 &	Y &	\\
18&	692&	2014-Apr-18&	12:55:00&	12:27:00&	12:31:00	&M7.3&	S17W34&	2036	&12	&Y$^{\dagger}$	&1203&	360&	Y&	j\\
19&	700&	2014-May-07$^{\ddagger}$&	16:30:00$^{\mathrm{b}}$	&16:03:00	&16:07:00&	M1.2	&S12W100&	2051&	-32	&N$^{\mathrm{d}}$&	923&	360&Y &\\	
20&	715&	2014-Jun-12&	22:00:00&	21:40:00$^{\mathrm{a}}$&	21:34:00&	M3.1&	S18W63&	2085&	0	&N$^{\mathrm{d}}$	&684&	186&	Y&\\	
21&	781&	2015-Jul-19&	\dots&	10:25:00$^{\mathrm{a}}$&	09:22:00&	C2.1	&S26W68&	2384	&9	&Y	&782&	194&	N&\\	
22&	789&	2015-Sep-20&	18:16:00&	18:00:00	&17:32:00&	M2.1&	S21W50&	2415	&-8&	Y$^{\dagger}$ &	1239	&360	&N&\\	
23&	790&	2015-Sep-30&	\dots&	13:16:00$^{\mathrm{a}}$	&13:18:00&	M1.1&	S20W55&	2422	&21	&Y$^{\dagger}$ &	536	&84&	Y&	j\\
\enddata

\tablecomments{
$^{\star}$Events measured by at least one of the two STEREO spacecraft. $^{\ddagger}$Events with backside source.
$^{\mathrm{a}}$~Wind, $^{\mathrm{b}}$~STEREO-A, and $^{\mathrm{c}}$~STEREO-B.
$^{\S}$~Flare started on the day prior to the proton event start day.
$\Delta\equiv L_{\mathrm{s}}-L_{\mathrm{f}}$, where $L_{\mathrm{s}}$ is the Stonyhurst longitude of the L1 (ACE, SOHO) magnetic connection point at $2.5\,R_{\odot}$ from the Sun center, and $L_{\mathrm{f}}$ is the Stonyhurst longitude of the flare.
$^{\mathrm{d}}$~The given AR does not exhibit open field lines to the ecliptic, but a nearby region within $\leq10^\circ$ in both latitude and longitude does show open field lines.
$^{\dagger}$~L1 connected to the AR.
$^{\dagger\dagger}$~New region not visible in the synoptic magnetic map.
}

\tablenotetext{i}{\citet{2014SoPh..289.3059R}.}
\tablenotetext{j}{\citet{2019ApJ...877...11C}.}
\tablenotetext{k}{\citet{2013ApJ...762...54W}.}
\tablenotetext{l}{\citet{2016ApJ...833...63B}.}
\tablenotetext{m}{\citet{2016ApJ...816...68D}.}
\tablenotetext{n}{\citet{2015ApJ...806..235N}.}

\end{deluxetable*}
\endgroup

Table~\ref{tab:source} summarizes the solar source characteristics of the identified proton events. Column 1 lists the event number, and Column 2 provides the corresponding ${}^3\mathrm{He}$-rich period number \citep{2022ApJS..263...22H}. Column 3 indicates the start date of the proton event in the 25--50\,MeV energy range. Fourteen of the events were observed by at least one of the two Solar TErrestrial RElations Observatory (STEREO) spacecraft using the High Energy Telescope \citep[HET;][]{2008SSRv..136..391V} in the 24--41\,MeV proton energy range. Six of these events (\#7, 9, 13, 14, 15, 17) were observed by both STEREO-A and STEREO-B. No HET measurements were available from either STEREO-A or STEREO-B for events \#22 and \#23, and STEREO-B HET data were also unavailable for event \#21. The minimum longitudinal separation between Earth (ACE, SOHO at L1) and the STEREO spacecraft was 70$^\circ$, indicating these 14 events were longitudinally widespread. Five events (\#7, 9, 13, 17, and 18) are also included in the NOAA list of Solar Proton Events Affecting the Earth’s Environment\footnote{\url{https://umbra.nascom.nasa.gov/SEP/}}. 

\begin{figure}
\epsscale{1.1}
\plotone{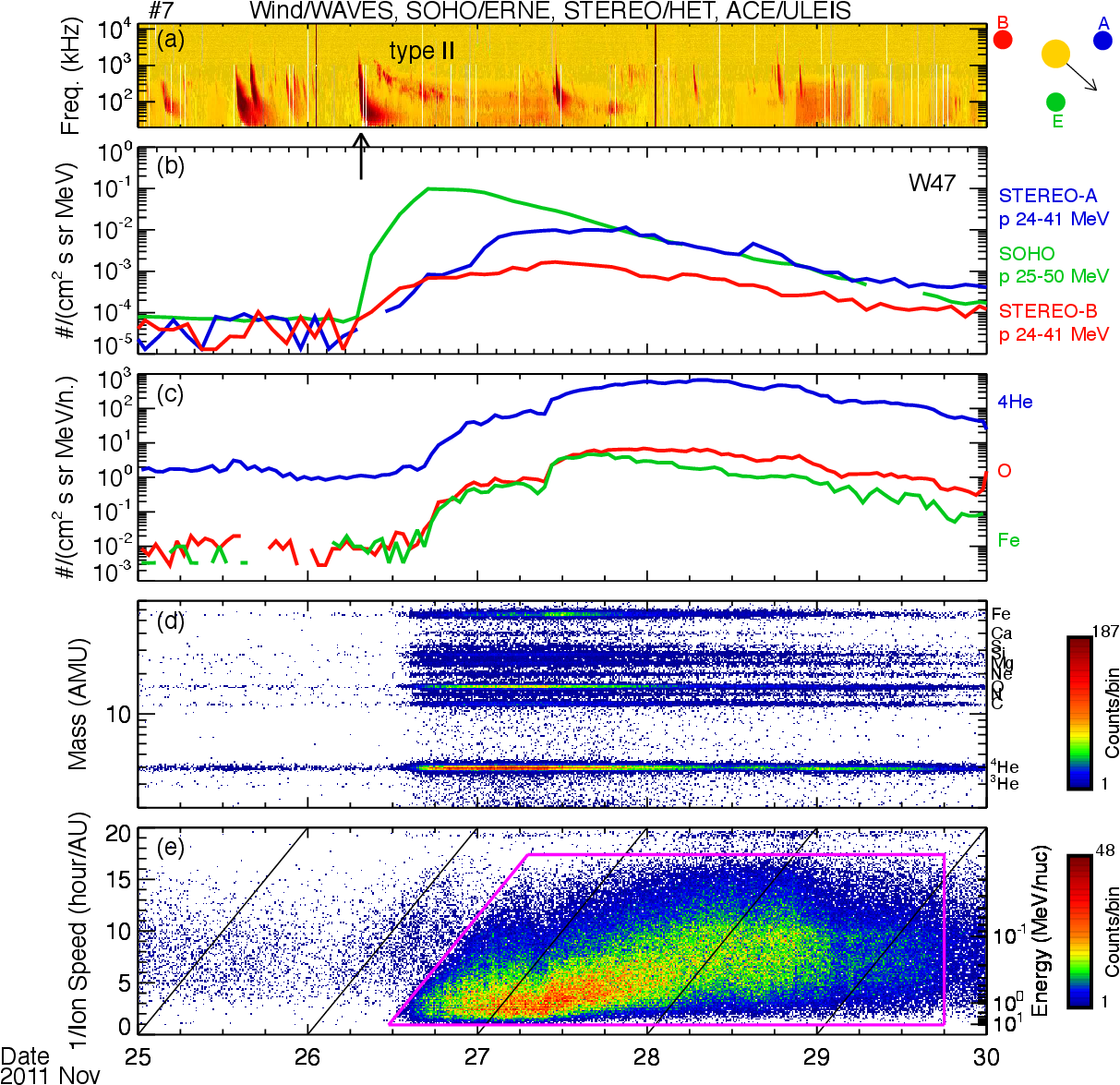}
\caption{Multi-spacecraft measurements of event \#7. (a) Wind/WAVES dynamic radio spectrogram. Color-coding indicates electric field intensity (dB above background). The vertical black arrow marks the type III radio burst associated with the event; the subsequent slow-drifting feature corresponds to a type II burst. The locations of STEREO-A (A), STEREO-B (B) and near-Earth (E) spacecraft (Wind, SOHO) at the time of the type III burst are shown in Heliocentric Earth Ecliptic (HEE) X-Y coordinates next to panel (a). The tilted arrow indicates the flare longitude (W47). (b) Two-hour proton intensities from STEREO-B/HET, STEREO-A/HET, and SOHO/ERNE in the indicated energy ranges. (c) One-hour ion intensities from ACE/ULEIS in 0.23--0.32\,MeV\,nucleon$^{-1}$ energy range. (d) Mass vs. time spectrogram from ACE/ULEIS covering 0.4--10\,MeV\,nucleon$^{-1}$. (e) Inverse ion speed vs. arrival time for ions with mass 10--70\,AMU. Slanted lines indicate expected arrival times for scatter-free propagation along a nominal Parker spiral path of 1.2\,au. The magenta polygon marks the region, defined in inverse speed-time space over which elemental abundance ratios were calculated.
\label{fig:panels}}
\end{figure}

Type II and type III radio bursts (Columns 4 and 5) were obtained from the NOAA Space Weather Prediction Center (SWPC) Edited Events list\footnote{\url{ftp://ftp.swpc.noaa.gov/pub/warehouse/}}. When not listed, we examined Wind/WAVES radio spectrograms \citep{1995SSRv...71..231B}, and when necessary, STEREO-A or STEREO-B WAVES data \citep{2008SSRv..136..487B}. No STEREO radio data were available for events \#21–23. Columns 6 and 7 provide the GOES X-ray flare start time (in UT, generally on the same day as the proton event onset unless otherwise noted) and the flare class, based on the SWPC catalog. Events \#2, 4, 11, 15, and 19 originated from the behind the limb. The flare associated with event \#13 began the day prior to the proton onset. Event \#14 was associated with multiple flares. Column 8 lists the flare location expressed in Stonyhurst heliographic coordinates: longitude (E/W) and latitude (S/N) relative to the solar central meridian and equator, respectively. The flare location was determined from the 131\,{\AA} extreme ultraviolet (EUV) channel of the Atmospheric Imaging Assembly \citep[AIA;][]{2012SoPh..275...17L} on SDO, which is sensitive to hot coronal plasma at temperatures of $\approx$10\,MK, which falls within the soft X-ray temperature range. For flares near or behind the limb, we used STEREO EUVI \citep{2008SSRv..136...67H} 171\,{\AA}  or 195\,{\AA}  images, selecting the channel with the higher cadence available. These channels are sensitive to cooler coronal temperatures of around 0.7\,MK and 1.5\,MK, respectively, while EUVI does not include channels sensitive to flare-like temperatures above 5\,MK. Column 9 gives the parent AR number. 

Column 10 shows the longitudinal separation between the flare site and the spacecraft’s magnetic footpoint at the source surface ($2.5\,R_{\odot}$, where magnetic fields are assumed radial), computed using the Parker spiral approximation with 1-hour solar wind speed data from the Coordinated Data Analysis Web\footnote{\url{https://cdaweb.gsfc.nasa.gov}}. Column 11 indicates whether open magnetic field lines from the parent AR were directed toward the ecliptic (intersect the source surface at latitudes $0^\circ$ and $\pm7^\circ$) and whether the spacecraft was magnetically connected to the AR via these field lines. Magnetic connectivity was evaluated using a potential field source surface (PFSS) model implemented in SolarSoft {\tt pfss} package\footnote{\url{https://www.lmsal.com/~derosa/pfsspack/}}. The model is sampled every 6\,hours and incorporates SDO HMI magnetograms assimilated into a flux-dispersal model to reconstruct the global photospheric magnetic field \citep{2003SoPh..212..165S}. 

Columns 12 and 13 list the associated CME speed (km\,s$^{-1}$) and angular width (degrees), based on the manually identified LASCO CME catalog\footnote{\url{https://cdaw.gsfc.nasa.gov/CME list/}} \citep{2004JGRA..109.7105Y}. No CME was reported for event \#23, likely due to observational obscuration by the background from two preceding CMEs (see Appendix~\ref{sec:cme23} where CME speed was estimated). Column 14 indicates the presence (‘Y’) or absence (‘N’) of a coronal jet in the parent AR, based on EUVI or AIA observations (Section~\ref{subsec:sources}). The final column cites prior studies that have reported these events. 

Figure~\ref{fig:panels} presents multi-spacecraft and multi-instrument observations of event \#7. As shown in Table~\ref{tab:source}, the event was associated with a relatively weak C1.2 X-ray flare and a halo CME with a speed of 933\,km\,s$^{-1}$. No EUV jet was observed in the source AR. Associated type III and type II radio bursts were clearly detected by Wind/WAVES (Fig.~\ref{fig:panels}a). The 2-hr high-energy proton intensity profiles in Fig.~\ref{fig:panels}b show that the event was also observed by STEREO-A and STEREO-B, which were separated by $\sim$106$^\circ$ in heliolongitude, from near-Earth spacecraft (ACE or SOHO). The positions of the relevant spacecraft are illustrated in Heliocentric Earth Ecliptic (HEE) coordinates adjacent to panel (a). Figure~\ref{fig:panels}c displays 1-hr He, O, and Fe intensities measured by ACE/ULEIS in the 0.23--0.32\,MeV\,nucleon$^{-1}$ energy range. The mass vs. time plot in Fig.~\ref{fig:panels}d, covering 0.4--10\,MeV\,nucleon$^{-1}$, shows a small concentration of ${}^3\mathrm{He}$ ions preceding the proton event. These ${}^3\mathrm{He}$ ions are at the end of a ${}^3\mathrm{He}$-rich period between 2011 November 22 21:00\,UT and November 26 12:00\,UT \citep{2022ApJS..263...22H}. A velocity-dispersive signature, where high-energy ions arrive earlier than lower-energy ones, is clearly visible in the inverse ion-speed vs. time spectrogram (Fig.~\ref{fig:panels}e). To reduce background from preceding and subsequent events, abundance ratios were computed within the polygonal region shown in Fig.~\ref{fig:panels}e. Numerical values corresponding to this sampling interval are provided in Table~\ref{tab:abunds}.

\subsection{Timing-based refinement}  \label{subsec:timing}

Guided by earlier studies that argue similar ion time profiles indicate a common acceleration and transport origin \citep{1999ApJ...525L.133M,2000AIPC..528..107W}, we compare, for each event in Table~\ref{tab:source}, the suprathermal ${}^3\mathrm{He}$ and ${}^4\mathrm{He}$ intensity profiles across the event window to identify and exclude cases contaminated by an intervening ISEP (or ${}^3\mathrm{He}$-rich) event. Five events (\#3, 15, 18, 21, and 22), showing distinct time profiles of ${}^3\mathrm{He}$ and ${}^4\mathrm{He}$, were caused by ${}^3\mathrm{He}$ from preceding ISEP events, and were excluded from the abundance analysis. A jet was observed in only one of these five proton events (\#18). Event \#3 was preceded by the 2011 January 27 ISEP event \citep{2015ApJ...806..235N,2021ApJ...908..243B}; event \#15 by the 2013 December 24 ISEP event \citep{2015ApJ...806..235N}; and event \#18 by the 2014 April 17 ISEP event \citep{2015ApJ...806..235N,2016ApJ...823..138M,2021ApJ...908..243B}. 

Four additional events (\#1, 5, 9, and 14) showed brief, distinct time profiles caused by independent ISEP events occurring during the proton events. These proton events were retained in the analysis by appropriately adjusting the integration intervals to exclude contamination. In event \#10, ${}^3\mathrm{He}$ and ${}^4\mathrm{He}$ intensities rose simultaneously; however, ${}^3\mathrm{He}$ exhibited only a brief enhancement, and no evidence of an independent source (e.g., a solar electron event or type III radio burst) was found. Thus, ${}^3\mathrm{He}$ in event \#10 was considered part of the main event. In total, 18 events remained, free from contamination by independent ISEP events. 

\begin{figure}
\centering
\epsscale{0.52}
\rotatebox{90}{\plotone{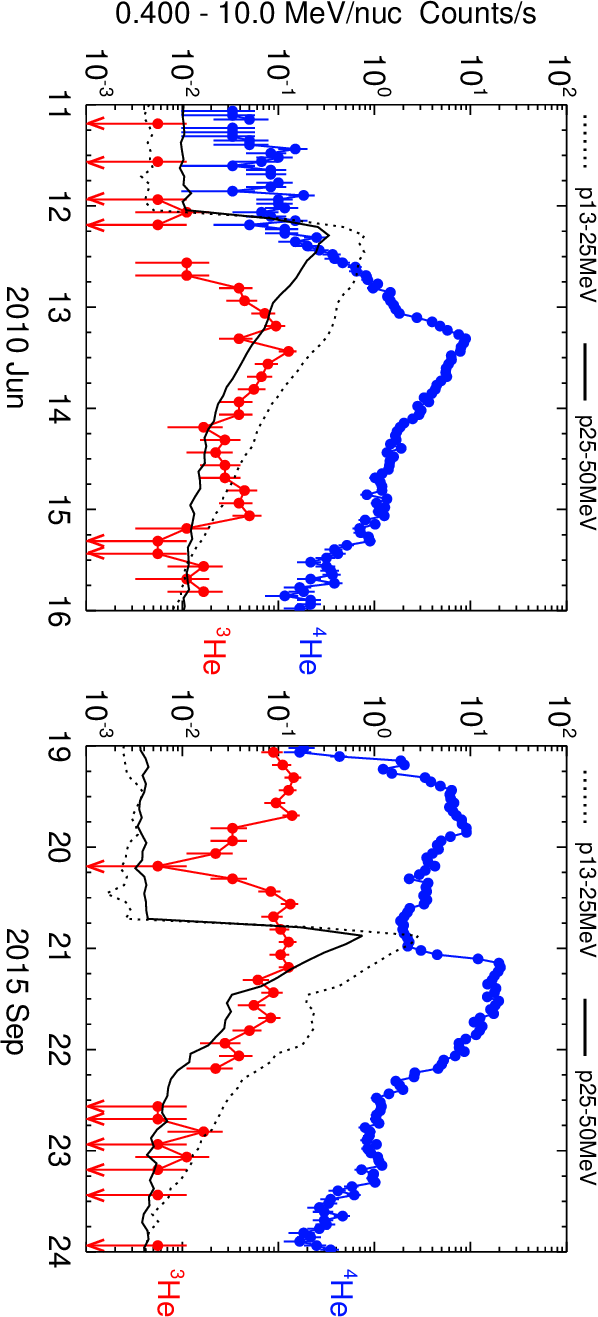}}
\caption{ACE/ULEIS 3-hr ${}^3\mathrm{He}$ (2.8--3.1\,AMU) and 1-hr ${}^4\mathrm{He}$ (3.5--6.0\,AMU) 0.4--10\,MeV\,nucleon$^{-1}$ count rates for events \#1 (left) and \#22 (right). Error bars with arrows denote ULEIS single-count data points. SOHO/ERNE 2-hr proton fluxes (in arbitrary units) in two energy ranges are overploted as black dotted and solid curves. 
\label{fig:counts}}
\end{figure}

Figure~\ref{fig:counts}(left) shows event \#1, where sporadic ${}^3\mathrm{He}$ counts were detected, prior to the main increase, possibly indicating residual ${}^3\mathrm{He}$ material. All 18 events exhibited similar precursors. The ${}^3\mathrm{He}$ and ${}^4\mathrm{He}$ profiles remained comparable until 2010 June 14, when a rise in ${}^3\mathrm{He}$ counts, associated with an independent ISEP event, was observed, accompanied by ACE/EPAM-detected solar energetic electrons \citep{1998SSRv...86..541G} and type III radio bursts at 00:47\,UT. A small bump was also visible in the ${}^4\mathrm{He}$ decay profile. In contrast, Figure~\ref{fig:counts}(right) presents event \#22 (excluded from the final list), where ${}^3\mathrm{He}$ counts continued to decline during the onset of the associated GSEP event on 2015 September 21. 

\section{Results} \label{sec:result}
\subsection{Elemental composition}  \label{subsec:element}

Elemental abundance ratios were determined for all 23 proton events included in this survey. The selected ratios are presented in Table~\ref{tab:abunds}. Columns 1 and 2 list the event number and year and Column 3 provides the ULEIS sampling intervals. Column 4 shows the ${}^3\mathrm{He}/{}^4\mathrm{He}$ ratio in the 0.5--2.0\,MeV\,nucleon$^{-1}$ energy range, while Columns 5--13 report heavy-ion elemental ratios in the 0.32--0.45\,MeV\,nucleon$^{-1}$ range. For the 14 events exhibiting clear velocity dispersion, the ratios were calculated within polygonal regions in the inverse ion-speed vs. time plots. Energetic storm particles were excluded from the elemental composition analysis. 

\begingroup
\setlength{\tabcolsep}{3pt}       \renewcommand{\arraystretch}{0.95} 

\begin{longrotatetable}
\begin{deluxetable}{cccccccccccccc}
\tabletypesize{\scriptsize}
\tablecaption{Elemental abundances for the proton events\label{tab:abunds}}
\tablewidth{\textheight} 
\startlongtable
\tablehead{
\colhead{\#} & \colhead{Year} & \colhead{Sampling interval} &
\colhead{${}^{3}\mathrm{He}/{}^{4}\mathrm{He}$} &
\colhead{${}^{4}\mathrm{He}/\mathrm{O}$} &
\colhead{$\mathrm{C}/\mathrm{O}$} & \colhead{$\mathrm{N}/\mathrm{O}$} &
\colhead{$\mathrm{Ne}/\mathrm{O}$} &
\colhead{$\mathrm{Mg}/\mathrm{O}$} & \colhead{$\mathrm{Si}/\mathrm{O}$} &
\colhead{$\mathrm{S}/\mathrm{O}$} & \colhead{$\mathrm{Ca}/\mathrm{O}$} &
\colhead{$\mathrm{Fe}/\mathrm{O}$} \\
\colhead{} & \colhead{} & \colhead{} &
\colhead{$(\times10^{2})$} & \colhead{} & \colhead{} & \colhead{} &
\colhead{} & \colhead{} & \colhead{} & \colhead{} & \colhead{} & \colhead{}
}
\decimals
\startdata
\finalevent{1} & 2010 & Jun 12, 18{:}00--Jun 14, 03{:}00$^{\mathrm{b}}$ &
1.54\,$\pm$\,0.18 & 408.2\,$\pm$\,25.5 & 0.320\,$\pm$\,0.040 & 0.120\,$\pm$\,0.022 & 0.337\,$\pm$\,0.041 &
0.337\,$\pm$\,0.041 & 0.310\,$\pm$\,0.041 & 0.065\,$\pm$\,0.017 & 0.070\,$\pm$\,0.018 & 0.928\,$\pm$\,0.086 \\
\finalevent{2} & 2010 & Sep 1, 03{:}00--Sep 1, 06{:}00 &
0.68\,$\pm$\,0.39 & 274.3\,$\pm$\,50.0 & 0.377\,$\pm$\,0.128 & 0.063\,$\pm$\,0.046 & 0.219\,$\pm$\,0.091 &
0.219\,$\pm$\,0.091 & 0.218\,$\pm$\,0.097 & 0.073\,$\pm$\,0.053 & \dots$^{\mathrm{c}}$ & 0.837\,$\pm$\,0.229 \\
3$^{\mathrm{a}}$ & 2011 & Jan 28, 18{:}00--Jan 30, 22{:}00 &
4.08\,$\pm$\,0.93 & 202.4\,$\pm$\,24.9 & 0.238\,$\pm$\,0.064 & 0.125\,$\pm$\,0.044 & 0.430\,$\pm$\,0.092 &
0.389\,$\pm$\,0.087 & 0.372\,$\pm$\,0.089 & 0.113\,$\pm$\,0.045 & 0.065\,$\pm$\,0.033 & 1.148\,$\pm$\,0.193 \\
4 & 2011&  Apr 21, 01{:}08--Apr 21, 19{:}00$^{\ast}$ &43.75\,$\pm$\,19.83 & 97.81\,$\pm$\,40.15 & 0.575\,$\pm$\,0.361 & \dots$^{\mathrm{c}}$ &
0.571\,$\pm$\,0.358 & 0.429\,$\pm$\,0.296 & 0.499\,$\pm$\,0.345 & \dots$^{\mathrm{d}}$ &
\dots$^{\mathrm{d}}$ & 0.499\,$\pm$\,0.344 \\
 & &Apr 21, 18{:}01--Apr 21, 19{:}00 & & & & & & & & & &\\
5 & 2011 & Aug 2, 12{:}04--Aug 4, 06{:}09$^{\ast,\mathrm{b}}$ & \dots & 177.6\,$\pm$\,6.5 & 0.344\,$\pm$\,0.024 & 0.094\,$\pm$\,0.011 &
0.159\,$\pm$\,0.015 & 0.218\,$\pm$\,0.018 & 0.244\,$\pm$\,0.020 & 0.067\,$\pm$\,0.010 &
0.008\,$\pm$\,0.004 & 0.240\,$\pm$\,0.019 \\
 & & Aug 3, 06{:}24--Aug 5, 01{:}48&  & & & & & & & & &\\
\finalevent{6} & 2011 &  Aug 8, 18{:}39--Aug 9, 08{:}38$^{\ast}$ &
1.05\,$\pm$\,0.09 & 129.9\,$\pm$\,3.0 & 0.282\,$\pm$\,0.013 & 0.106\,$\pm$\,0.007 &
0.233\,$\pm$\,0.011 & 0.396\,$\pm$\,0.016 & 0.504\,$\pm$\,0.020 & 0.111\,$\pm$\,0.008 &
0.079\,$\pm$\,0.016 & 1.085\,$\pm$\,0.030 \\
& &  Aug 9, 14{:}18--Aug 10, 04{:}18& & & & & & & & & &\\
\finalevent{7} & 2011 & Nov 26, 10{:}44--Nov 29, 18{:}00$^{\ast}$ & 0.32\,$\pm$\,0.03 & 97.02\,$\pm$\,1.07 & 0.313\,$\pm$\,0.006 & 0.100\,$\pm$\,0.003 &
0.114\,$\pm$\,0.004 & 0.199\,$\pm$\,0.005 & 0.261\,$\pm$\,0.006 & 0.065\,$\pm$\,0.003 &
0.022\,$\pm$\,0.003 & 0.458\,$\pm$\,0.007 \\
 & & Nov 27, 08{:}02--Nov 29, 18{:}00& & & & & & & & & &\\
\finalevent{8} & 2012 & Jan 20, 03{:}00--Jan 22, 00{:}00 &
0.99\,$\pm$\,0.08 & 138.3\,$\pm$\,3.0 & 0.300\,$\pm$\,0.013 & 0.117\,$\pm$\,0.007 &
0.123\,$\pm$\,0.008 & 0.185\,$\pm$\,0.010 & 0.201\,$\pm$\,0.011 & 0.038\,$\pm$\,0.004 &
0.017\,$\pm$\,0.006 & 0.227\,$\pm$\,0.011 \\
\finalevent{9} & 2012 & May 17, 04{:}06--May 19, 12{:}00$^{\ast,\mathrm{b}}$ &0.35\,$\pm$\,0.04 & 91.59\,$\pm$\,1.23 & 0.337\,$\pm$\,0.008 & 0.107\,$\pm$\,0.004 &
0.145\,$\pm$\,0.005 & 0.194\,$\pm$\,0.006 & 0.211\,$\pm$\,0.006 & 0.056\,$\pm$\,0.003 &
0.019\,$\pm$\,0.003 & 0.371\,$\pm$\,0.006 \\
 & & May 18, 07{:}58--May 19, 12{:}00& & & & & & & & & &\\
\finalevent{10} & 2012 & Jun 14, 19{:}08--Jun 16, 09{:}00$^{\ast}$ &
0.29\,$\pm$\,0.06 & 105.6\,$\pm$\,1.6 & 0.352\,$\pm$\,0.009 & 0.121\,$\pm$\,0.005 &
0.151\,$\pm$\,0.006 & 0.159\,$\pm$\,0.006 & 0.125\,$\pm$\,0.005 & 0.027\,$\pm$\,0.002 &
0.005\,$\pm$\,0.002 & 0.047\,$\pm$\,0.002\\
& &  Jun 15, 14{:}48--Jun 16, 09{:}00& & & & & & & & & &\\
\finalevent{11} & 2013 & Feb 26, 20{:}00--Feb 28, 00{:}00 &
0.59\,$\pm$\,0.19 & 132.8\,$\pm$\,8.7 & 0.358\,$\pm$\,0.043 & 0.095\,$\pm$\,0.020 &
0.193\,$\pm$\,0.030 & 0.148\,$\pm$\,0.025 & 0.159\,$\pm$\,0.028 & 0.048\,$\pm$\,0.015 &
\dots$^{\mathrm{d}}$ & 0.084\,$\pm$\,0.020 \\
12 & 2013 & May 2, 18{:}00--May 4, 15{:}00 &
20.05\,$\pm$\,2.57 & 142.3\,$\pm$\,18.7 & 0.217\,$\pm$\,0.064 & 0.123\,$\pm$\,0.046 &
0.231\,$\pm$\,0.066 & 0.338\,$\pm$\,0.083 & 0.466\,$\pm$\,0.108 &0.161\,$\pm$\,0.057 & \dots$^{\mathrm{d}}$ & 0.913\,$\pm$\,0.171 \\
\finalevent{13} & 2013 & Sep 30, 03{:}26--Oct 1, 22{:}00$^{\ast}$ &
0.36\,$\pm$\,0.04 & 91.57\,$\pm$\,1.59 & 0.344\,$\pm$\,0.011 & 0.108\,$\pm$\,0.005 &
0.122\,$\pm$\,0.006 & 0.201\,$\pm$\,0.008 & 0.260\,$\pm$\,0.009 & 0.080\,$\pm$\,0.005 &
0.025\,$\pm$\,0.004 & 0.498\,$\pm$\,0.010 \\
& &  Oct 1, 04{:}00--Oct 1, 22{:}00& & & & & & & & & &\\
\finalevent{14} & 2013 & Oct 25, 14{:}51--Oct 27, 13{:}00$^{\ast,\mathrm{b}}$ &
2.45\,$\pm$\,0.33 & 100.8\,$\pm$\,5.1 & 0.254\,$\pm$\,0.026 & 0.106\,$\pm$\,0.016 &
0.156\,$\pm$\,0.020 & 0.225\,$\pm$\,0.024 & 0.229\,$\pm$\,0.026 & 0.056\,$\pm$\,0.012 &
0.018\,$\pm$\,0.007 & 0.303\,$\pm$\,0.031 \\
& &  Oct 27, 09{:}01--Oct 27, 13{:}00 & & & & & & & & & &\\
15$^{\mathrm{a}}$ & 2013 & Dec 27, 12{:}00--Dec 28, 22{:}00 &
2.28\,$\pm$\,0.56 & 122.1\,$\pm$\,9.2 & 0.262\,$\pm$\,0.041 & 0.130\,$\pm$\,0.027 &
0.120\,$\pm$\,0.026 & 0.245\,$\pm$\,0.039 & 0.146\,$\pm$\,0.031 & 0.070\,$\pm$\,0.021 &
0.023\,$\pm$\,0.012 & 0.232\,$\pm$\,0.040 \\
\finalevent{16} & 2014 & Jan 5, 04{:}08--Jan 6, 10{:}49$^{\ast}$ &
0.29\,$\pm$\,0.08 & 114.3\,$\pm$\,3.1 & 0.268\,$\pm$\,0.015 & 0.118\,$\pm$\,0.009 &
0.138\,$\pm$\,0.010 & 0.188\,$\pm$\,0.012 & 0.217\,$\pm$\,0.014 & 0.049\,$\pm$\,0.006 &
0.016\,$\pm$\,0.006 & 0.212\,$\pm$\,0.013 \\
& &  Jan 5, 23{:}48--Jan 7, 09{:}46 & & & & & & & & & &\\
\finalevent{17} & 2014 & Feb 25, 18{:}00--Mar 2, 21{:}00 &
0.30\,$\pm$\,0.03 & 103.1\,$\pm$\,0.9 & 0.362\,$\pm$\,0.006 & 0.109\,$\pm$\,0.003 &
0.113\,$\pm$\,0.003 & 0.157\,$\pm$\,0.004 & 0.147\,$\pm$\,0.004 & 0.031\,$\pm$\,0.002 &
0.008\,$\pm$\,0.002 & 0.116\,$\pm$\,0.003 \\
18$^{\mathrm{a}}$ & 2014 & Apr 18, 15{:}39--Apr 20, 06{:}00$^{\ast}$ &
0.69\,$\pm$\,0.06 & 103.0\,$\pm$\,1.5 & 0.298\,$\pm$\,0.008 & 0.115\,$\pm$\,0.005 &
0.194\,$\pm$\,0.006 & 0.269\,$\pm$\,0.007 & 0.261\,$\pm$\,0.008 & 0.060\,$\pm$\,0.003 &
0.027\,$\pm$\,0.004 & 0.594\,$\pm$\,0.011 \\
& &  Apr 19, 11{:}18--Apr 20, 06{:}00 & & & & & & & & & &\\
\finalevent{19} & 2014 & May 8, 06{:}00--May 9, 06{:}00 &
1.96\,$\pm$\,0.57 & 137.0\,$\pm$\,22.0 & 0.183\,$\pm$\,0.070 & 0.091\,$\pm$\,0.048 &
0.159\,$\pm$\,0.065 & 0.250\,$\pm$\,0.084 & 0.238\,$\pm$\,0.087 & 0.053\,$\pm$\,0.038 &
\dots$^{\mathrm{d}}$ & 0.979\,$\pm$\,0.218 \\
\finalevent{20} & 2014 & Jun 12, 23{:}43--Jun 15, 18{:}00$^{\ast}$ &
4.20\,$\pm$\,0.30 & 152.4\,$\pm$\,4.3 & 0.252\,$\pm$\,0.015 & 0.090\,$\pm$\,0.008 &
0.231\,$\pm$\,0.014 & 0.308\,$\pm$\,0.017 & 0.309\,$\pm$\,0.018 & 0.076\,$\pm$\,0.008 &
0.059\,$\pm$\,0.018 & 0.809\,$\pm$\,0.034 \\
& &  Jun 14, 05{:}12--Jun 15, 18{:}00  & & & & & & & & & &\\
21$^{\mathrm{a}}$ & 2015 & Jul 19, 17{:}41--Jul 20, 22{:}31$^{\ast}$ &
0.35\,$\pm$\,0.27 & 103.1\,$\pm$\,4.6 & 0.300\,$\pm$\,0.026 & 0.113\,$\pm$\,0.015 &
0.107\,$\pm$\,0.014 & 0.171\,$\pm$\,0.019 & 0.163\,$\pm$\,0.019 & 0.043\,$\pm$\,0.009 &
0.006\,$\pm$\,0.003 & 0.130\,$\pm$\,0.016 \\
& &  Jul 20, 14{:}10--Jul 20, 22{:}31  & & & & & & & & & &\\
22$^{\mathrm{a}}$ & 2015 & Sep 20, 19{:}32--Sep 22, 12{:}00$^{\ast}$ &
0.49\,$\pm$\,0.07 & 51.43\,$\pm$\,0.71 & 0.235\,$\pm$\,0.006 & 0.097\,$\pm$\,0.004 &
0.150\,$\pm$\,0.005 & 0.191\,$\pm$\,0.006 & 0.174\,$\pm$\,0.006 & 0.030\,$\pm$\,0.002 &
0.007\,$\pm$\,0.002 & 0.056\,$\pm$\,0.002 \\
& &  Sep 21, 13{:}34--Sep 22, 12{:}00 & & & & & & & & & &\\
\finalevent{23} & 2015 & Sep 30, 21{:}56--Oct 03, 18{:}00$^{\ast}$ &
0.58\,$\pm$\,0.06 & 138.1\,$\pm$\,2.3 & 0.281\,$\pm$\,0.009 & 0.091\,$\pm$\,0.005 &
0.131\,$\pm$\,0.006 & 0.183\,$\pm$\,0.007 & 0.212\,$\pm$\,0.008 & 0.055\,$\pm$\,0.004 &
0.020\,$\pm$\,0.005 & 0.364\,$\pm$\,0.011 \\
& &  Oct 01, 22{:}31--Oct 03, 18{:}00 & & & & & & & & & &\\
\enddata
\tablecomments{
    ${}^{3}$He/${}^{4}$He at 0.5--2.0\,MeV\,nucleon$^{-1}$; all other ratios at 0.32--0.45\,MeV\,nucleon$^{-1}$. $^{\mathrm{a}}$~Event includes $^{3}$He originating from a preceding impulsive event.
    $^{\mathrm{b}}$~Integration intervals adjusted to exclude $^{3}$He contribution from impulsive events occurring during the proton event.
    $^{\ast}$Polygon integration intervals--upper and lower rows correspond to 10\,MeV\,nucleon$^{-1}$ and 0.03\,MeV\,nucleon$^{-1}$, respectively.
    $^{\mathrm{c}}$~Single count detected for N, Ca. $^{\mathrm{d}}$~No counts detected for S, Ca.
    Boldface event numbers denote the final sample of 15 (see text for abundance-based exclusion criteria).
  }\end{deluxetable}
\end{longrotatetable}

\endgroup

\begin{figure}
\centering
\epsscale{0.56}
\rotatebox{90}{\plotone{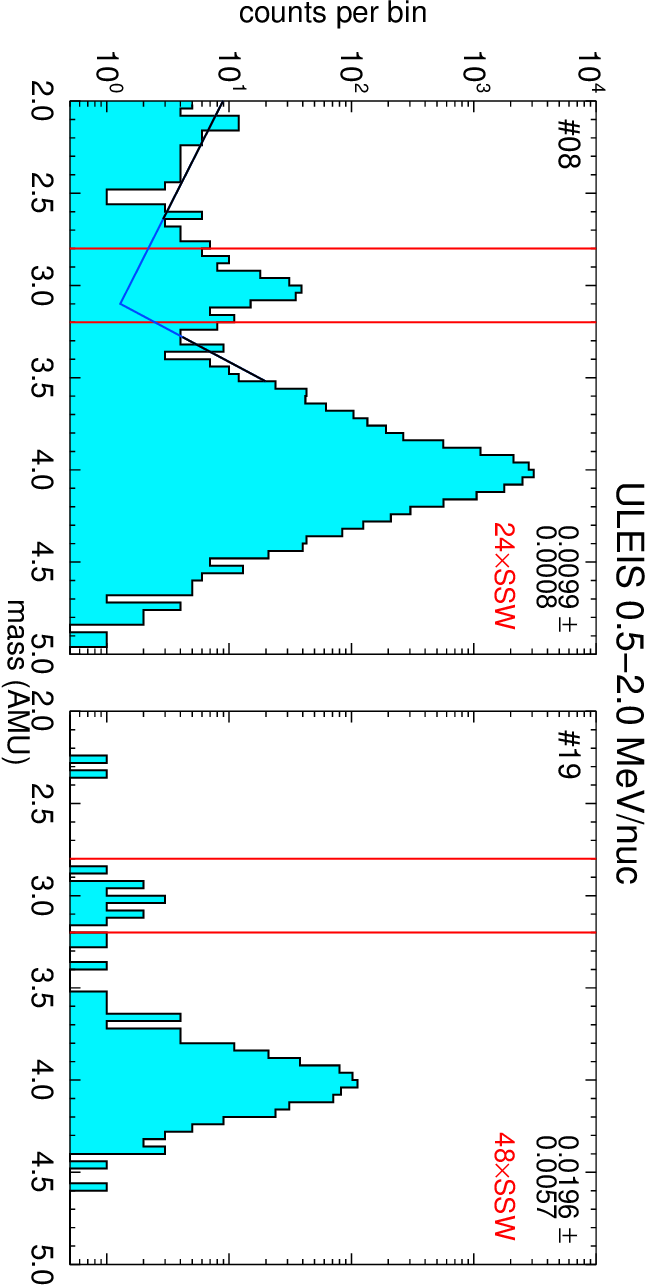}}
\caption{Helium mass histograms for events \#8 (left) and \#19 (right). The vertical red lines mark the ${}^3\mathrm{He}$ mass range. The ${}^3\mathrm{He}/{}^4\mathrm{He}$ ratios are shown in the upper right corner. Left: The two tilted black lines are the least square fits of the background and  ${}^4\mathrm{He}$ spillover. The extrapolations to the ${}^3\mathrm{He}$ range are marked by blue.
\label{fig:hist}}
\end{figure}

The ${}^3\mathrm{He}/{}^4\mathrm{He}$ ratio was calculated from helium (2--5\,AMU) mass histograms. A relatively broad energy range (0.5--2.0\,MeV\,nucleon$^{-1}$) was used to ensure sufficient ${}^3\mathrm{He}$ counts. Background (2.00--2.64\,AMU) and ${}^4\mathrm{He}$ spillover (3.28--3.52\,AMU) were modeled using two separate logarithmic fits of the form $\log(y) = a + bx$, where $x$ is the ion mass and $y$ the counts. These fits were extrapolated into the ${}^3\mathrm{He}$ mass range (2.8--3.2\,AMU), and the estimated background counts were subtracted from the total counts in this interval. Net ${}^4\mathrm{He}$ counts were obtained from the 3.5--5.0\,AMU range. Figure~\ref{fig:hist} illustrates this process for two events: event \#8, where background and spillover subtraction was necessary, and event \#19, where no significant background or spillover was present. For event \#5, adjusting the integration interval to exclude independent ${}^3\mathrm{He}$-rich event resulted in zero net ${}^3\mathrm{He}$ counts. For the remaining events, the net ${}^3\mathrm{He}$ counts exceeded zero by at least 2$\sigma$, consistent with the detection threshold used in \citet{2006ApJ...649..470D}. Events \#4 and \#12 exhibited particularly high ${}^3\mathrm{He}/{}^4\mathrm{He}$ ratios.

\begin{figure}
\centering
\epsscale{1.1}
\plotone{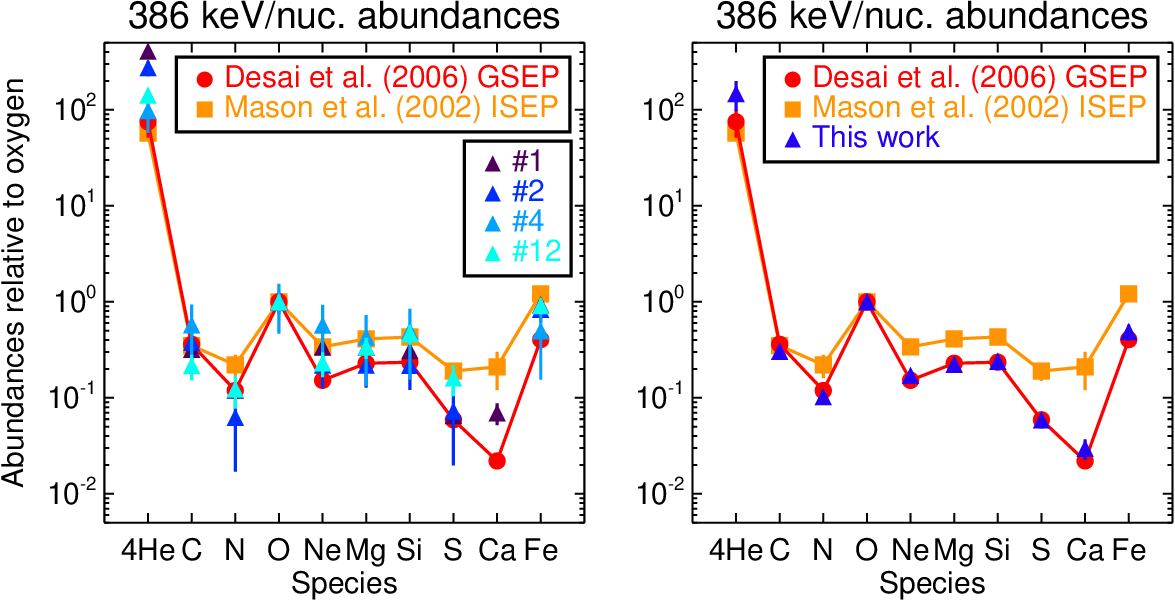}
\caption{Left: Element abundances relative to O for events \#1, 2, 4, 12 in this study (blue-shaded triangles). Right: Survey-averaged elemental abundances relative to O (blue triangles). For comparison, GSEP and ISEP reference abundances are shown as red circles and orange squares, respectively. 
\label{fig:abu}}
\end{figure}

Figure~\ref{fig:abu}(left) shows the relative abundances for events \#1, 2, 4, and 12 along with GSEP \citep{2006ApJ...649..470D} and ISEP \citep{2002ApJ...574.1039M} average abundances at the same low energies. We focus on elements from Ne to Fe, where the differences between GSEP and ISEP reference abundances are most pronounced. Event \#12 appears in the ISEP event list compiled by \citet{2015ApJ...806..235N}. The abundances of Si, S, and Fe in event \#12 are ISEP-like, while Ne and Mg are enhanced relative to GSEP values. Ca was not measured. This high-energy proton event was accompanied by a type II radio burst and a partial halo CME. Event \#4, which exhibited the highest ${}^3\mathrm{He}/{}^4\mathrm{He}$ ratio in the survey, showed ISEP-like abundances for Ne, Mg, and Si, but not for Fe. Abundances of S and Ca were not obtained. Similar to event \#12, event \#4 was accompanied by a type II radio burst and a partial halo CME. Additionally, the event \#4 was widespread, observed also by STEREO-A, which was separated by 91$^\circ$ in heliolongitude from L1. These two events are anomalous in their characteristics and may not represent typical ISEP events with only a single (flare) component.

Events \#1 and \#2 were included in the ISEP event list by \citet{2013ApJ...762...54W} and \citet{2016ApJ...833...63B}. These events exhibited ${}^3\mathrm{He}/{}^4\mathrm{He}$ ratios of 38$\times$SSW and 17$\times$SSW, respectively, values typical of  ${}^3\mathrm{He}$-enriched GSEP events. With the exception of Fe (and Ne in event \#1), the heavy-ion composition in both events were GSEP-like. The Ca abundance is unknown for event \#2.  Both events were widespread, observed by STEREO-A at separation of 74$^\circ$ and 81$^\circ$ in heliolongitude from L1, associated with wide CMEs (including a halo CME in event \#2) and type II radio bursts, indicative of a shock acceleration. These two high-energy proton events were very likely GSEP events, enriched in ${}^3\mathrm{He}$ and Fe.

Figure~\ref{fig:abu}(right) presents the survey-averaged, unweighted abundances using the final set of 15 events listed in Table~\ref{tab:abunds} (\#1, 2, 6--11, 13, 14, 16, 17, 19, 20, and 23). Event \#5, which showed no detectable ${}^3\mathrm{He}$, and events \#4 and \#12, which exhibited extreme ${}^3\mathrm{He}$ enrichments, were excluded from the averaging. Notably, event \#5 did not show a jet in the source, whereas events \#4 and \#12 did. The figure demonstrates that the relative abundances measured in this study are consistent with the GSEP reference values at the same low energies. Table~\ref{tab:sepab} summarizes the average heavy-ion abundances from this survey, along with GSEP \citep{2006ApJ...649..470D} and ISEP \citep{2002ApJ...574.1039M} averages. Following the methodology of \citet{2006ApJ...649..470D}, the uncertainties in the survey-averaged abundances presented in Table~\ref{tab:sepab} are computed by combining the standard error of the mean in quadrature with estimated systematic uncertainties of 31\% for ${}^4\mathrm{He}$ and 2\% for C--Fe.

\begin{deluxetable}{cccc}
\tabletypesize{\scriptsize}
\tablecaption{Average SEP heavy ion abundances\label{tab:sepab}}
\tablewidth{0pt}
\tablehead{
\colhead{Element} &
\colhead{GSEP$^{\mathrm{a}}$} &
\colhead{GSEP$^{\mathrm{b}}$} &
\colhead{ISEP$^{\mathrm{c}}$} \\
\colhead{} &
\colhead{(386 keV\,nuc.$^{-1}$)} &
\colhead{(385 keV\,nuc.$^{-1}$)} &
\colhead{(385 keV\,nuc.$^{-1}$)}
}
\startdata
$^{4}$He & $147.7 \pm 50.7$ & $75.0 \pm 23.6$ & $57 \pm 7.79$ \\
C        & $0.306 \pm 0.015$ & $0.361 \pm 0.012$ & $0.35 \pm 0.08$ \\
N        & $0.103 \pm 0.004$ & $0.119 \pm 0.003$ & $0.22 \pm 0.06$ \\
O        & $\equiv 1.0 \pm 0.02$ & $\equiv 1.0 \pm 0.02$ & $\equiv 1.0 \pm 0.14$ \\
Ne       & $0.171 \pm 0.016$ & $0.152 \pm 0.005$ & $0.34 \pm 0.06$ \\
Mg       & $0.223 \pm 0.019$ & $0.229 \pm 0.007$ & $0.41 \pm 0.07$ \\
Si       & $0.240 \pm 0.024$ & $0.235 \pm 0.011$ & $0.43 \pm 0.06$ \\
S        & $0.059 \pm 0.006$ & $0.059 \pm 0.004$ & $0.19 \pm 0.04$ \\
Ca       & $0.030 \pm 0.007$ & $0.022 \pm 0.002$ & $0.21 \pm 0.09$ \\
Fe       & $0.488 \pm 0.091$ & $0.404 \pm 0.047$ & $1.21 \pm 0.14$ \\
\hline
\multicolumn{1}{c}{Ratio} & (0.5--2.0 MeV\,nuc.$^{-1}$) & (0.5--2.0 MeV\,nuc.$^{-1}$) & (385 keV\,nuc.$^{-1}$) \\
$^{3}$He/$^{4}$He & $0.011 \pm 0.003$ & $0.006 \pm 0.002$ & $0.354 \pm 0.137$ \\
\enddata
\tablecomments{$^{\mathrm{a}}$This work. $^{\mathrm{b}}$From \citet{2006ApJ...649..470D}. $^{\mathrm{c}}$From \citet{2002ApJ...574.1039M} .}
\end{deluxetable}

\begin{figure}
\centering
\epsscale{1.1}
\plotone{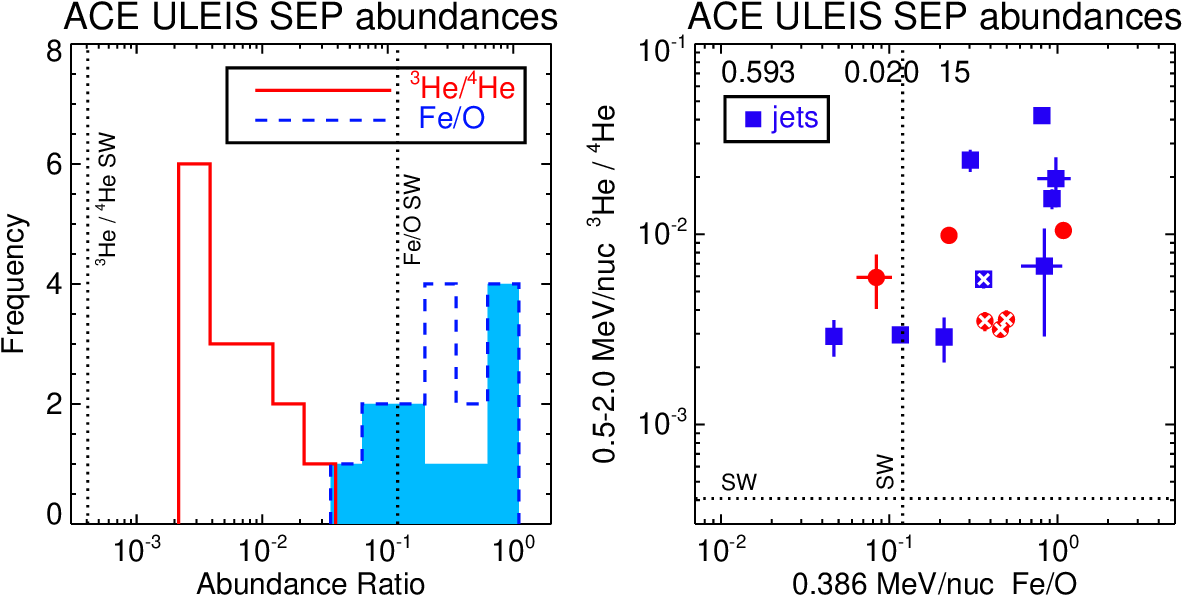}
\caption{Left: Distributions of 0.5--2.0\,MeV\,nucleon$^{-1}$ ${}^3\mathrm{He}/{}^4\mathrm{He}$ (red solid line) and 386\,keV\,nucleon$^{-1}$ Fe/O (blue dashed line). The two vertical dotted lines indicate the slow solar wind ${}^3\mathrm{He}/{}^4\mathrm{He}$ ratio \citep{1998SSRv...84..275G} and Fe/O ratio \citep{2000JGR...10527217V}. Right: Scatter plot of ${}^3\mathrm{He}/{}^4\mathrm{He}$ versus Fe/O. The three numbers indicate the Spearman correlation coefficient, the $p$-value, and the sample size. Note that outlier event \#4 (${}^3\mathrm{He}/{}^4\mathrm{He} \sim4 \times 10^{-1}$) and event \#12 (${}^3\mathrm{He}/{}^4\mathrm{He} \sim 2 \times 10^{-1}$) are not considered. Blue squares mark events associated with jets at the source; red circles denote events without observed jets. White crosses represent events with Fe and O time profiles that differ from the remaining events (see text). The blue-shaded histogram (left panel) excludes these events. Dotted horizontal and vertical lines indicate slow solar wind values.
\label{fig:hist-abu}}
\end{figure}

A histogram and scatter plot of ${}^3\mathrm{He}/{}^4\mathrm{He}$ and Fe/O ratios are shown in Figure~\ref{fig:hist-abu}. The left panel indicates distinct distributions of the ${}^3\mathrm{He}/{}^4\mathrm{He}$ and Fe/O ratios. While ${}^3\mathrm{He}/{}^4\mathrm{He}$ distribution peaks at the lowest values, Fe/O distribution peaks at higher values (dashed line). The 0.23--0.32\,MeV\,nucleon$^{-1}$ O and Fe intensity time profiles display distinct behavior in four western events (\#7, 9, 13, and 23): during the rising phase of the intensity profile, Fe closely tracks O, but diverges during the decay phase. The enhanced Fe/O observed during the injection phase of these events is likely a consequence of magnetic-rigidity-dependent transport effects \citep[e.g.,][]{1999ApJ...525L.133M}. These events show similar event-integrated Fe/O values, close to the survey average, and relatively low ${}^3\mathrm{He}/{}^4\mathrm{He}$ ratios (see Fig.~\ref{fig:hist-abu}, right panel, where these events are marked by white crosses). These four events are excluded in the shaded blue histogram in Fig.~\ref{fig:hist-abu}(left panel), resulting in a more uniform distribution. 

The minimum, maximum, mean, and median ${}^3\mathrm{He}/{}^4\mathrm{He}$ ratios, expressed as multiples of the SSW reference value ($4.08 \pm 0.25)\times10^{-4}$, are 7, 103, $26 \pm 7$, and 15, respectively. For comparison, \citet{2006ApJ...649..470D} found that 14 of 64 GSEP events ($\sim$22\%) exhibited ${}^3\mathrm{He}/{}^4\mathrm{He}\geq7$ times the SSW value (i.e., the minimum enhancement observed in this survey), while \citet{2016ApJ...816...68D} reported 16 of 46 events ($\sim$35\%) meeting this criterion. As previously mentioned, events \#4 and \#12 exhibit extreme enrichments, with ${}^3\mathrm{He}/{}^4\mathrm{He}$ ratios of 502 and 1072 times the SSW value, respectively, both exceeding ${}^3\mathrm{He}$ enhancements reported in earlier GSEP studies.

Figure~\ref{fig:hist-abu}(right panel) reveals a moderate positive correlation between ${}^3\mathrm{He}/{}^4\mathrm{He}$ and Fe/O, with $r = 0.593$ and a low probability ($p = 2$\%) that the correlation arose by chance. This trend is primarily driven by events associated with jets, for which $r = 0.633$ and $p = 7$\%.

Figure~\ref{fig:hist-abu} shows that ${}^3\mathrm{He}$-enriched GSEP events in this survey exhibit Fe/O ratios that are enhanced to ISEP-like values ($1.21\pm0.14$; see Table~\ref{tab:sepab}), likely influenced by transport, or close to the slow solar wind Fe/O ratio of $0.120 \pm 0.024$ \citep{2000JGR...10527217V}.

\begin{figure}
\centering
\epsscale{0.55}
\plotone{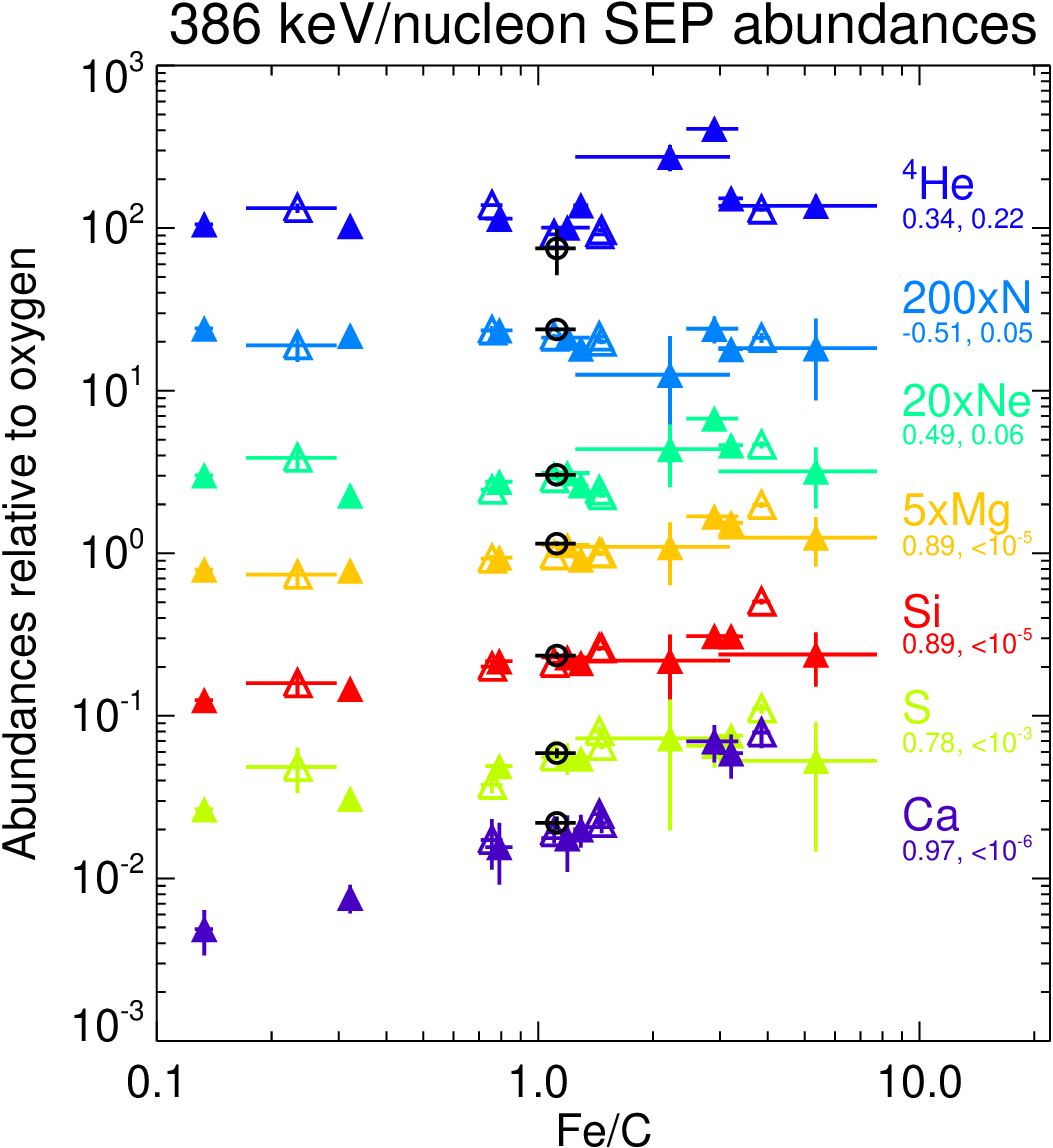}
\caption{Element abundance ratios relative to oxygen (O) plotted against Fe/C for the final set of 15 events listed in Table~\ref{tab:abunds}. Events associated with jets are shown as solid triangles; those without jets are shown as open triangles. The two numbers below each element label represent the Spearman correlation coefficient and the corresponding $p$-value. For clarity, some abundance ratios have been scaled by the multiplicative factor indicated. Open black circles mark the corresponding average GSEP abundances from \citet{2006ApJ...649..470D}.
\label{fig:heavy}}
\end{figure}

Figure~\ref{fig:heavy} presents the heavy-ion abundance enhancement pattern derived from Table~\ref{tab:abunds} for the final set of 15 ${}^3\mathrm{He}$-enriched GSEP events included in this survey. Enhancements in the Ne/O, Mg/O, Si/O, S/O and Ca/O ratios are consistently accompanied by simultaneous increases in the Fe/C ratio. A similar behavior for S/O and Ca/O was reported by \citet{2006ApJ...649..470D} in a broader survey of GSEP events, although fewer than half of those events were ${}^3\mathrm{He}$-enriched. In contrast, the N/O ratio exhibits an inverse trend, while the ${}^4\mathrm{He}$/O ratio displays only a weak positive correlation with Fe/C. Except for Ca/O, the event-to-event variations in the abundance ratios shown in Fig.~\ref{fig:heavy} are typically less than an order of magnitude. The Ca/O and Fe/C, as well as Fe/O and  ${}^3\mathrm{He}/{}^4\mathrm{He}$  (see Fig.~\ref{fig:hist-abu}), shows larger variations, reaching up to $\sim1.5$ orders of magnitude. Furthermore, Fig.~\ref{fig:heavy} indicates that the abundances in events without jets (open triangles) are not systematically different from those in events associated with jets (solid triangles).

\subsection{Magnetic connection}  \label{subsec:connect}

\begin{figure}
\centering
\epsscale{1.1}
\plotone{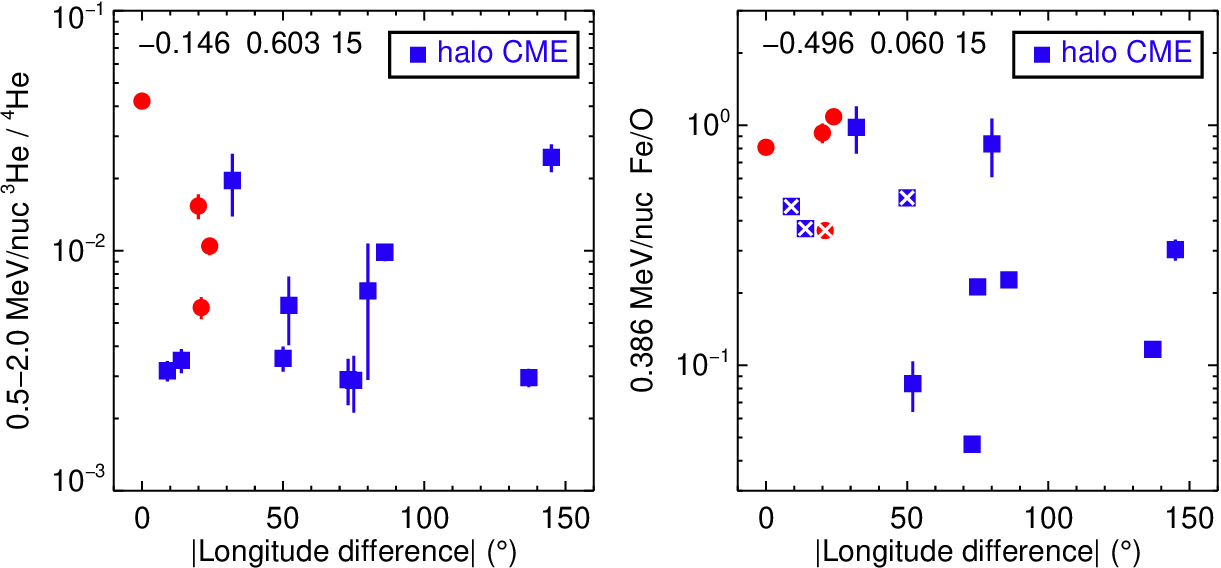}
\caption{${}^3\mathrm{He}/{}^4\mathrm{He}$ and Fe/O versus the absolute value of the longitudinal separation between the ACE magnetic footpoint at $2.5\,R_{\odot}$ from the Sun center and the flare location. Blue squares indicate events associated with halo CMEs, and red circles indicate events associated with partial halo CMEs. Points marked with white crosses (right panel), as well as the the three numerical values shown,  have the same meaning as in Fig.~\ref{fig:hist-abu}.
\label{fig:halo}}
\end{figure}

Figure~\ref{fig:halo}(left) shows no apparent correlation between ${}^3\mathrm{He}/{}^4\mathrm{He}$ and the longitudinal separation between the ACE magnetic footpoint at the Sun and the associated flare. In contrast, Figure~\ref{fig:halo}(right) displays a moderate negative correlation between Fe/O and this longitudinal difference. Although the number of events is small, the contrasting behaviors of ${}^3\mathrm{He}/{}^4\mathrm{He}$ , and Fe/O are evident. Removing the event at the largest longitudinal difference (\#14; also, among the highest ${}^3\mathrm{He}/{}^4\mathrm{He}$) changes the correlation coefficient to $-0.349$ ($p=0.221$) for ${}^3\mathrm{He}/{}^4\mathrm{He}$ and to $-0.459$ ($p=0.098$) for Fe/O. Figure~\ref{fig:halo} also shows that events with an absolute longitudinal separation of $< 30^\circ$ are associated with (narrower) partial-halo CMEs, whereas those with absolute longitudinal separation between 30$^\circ$ and 150$^\circ$ are more often associated with halo CMEs. This is consistent with wide CMEs intersecting a broader range of longitudes.

Using a PFSS coronal field model, we find that in 8 of the 15 events the parent AR shows open field lines to the ecliptic. The Parker-spiral approximation indicates that in two of these eight the modeled field lines connect the spacecraft directly to the AR. Field line meandering can produce spreads of $\sim10^\circ$ at 1\,au in both longitude and latitude \citep{2019ApJ...887..102M}. Accordingly, nominal Parker-spiral connections may be altered, and even without open field lines to the ecliptic, a connection to L1 can still occur. In a further four of the 15 events, open field lines to the ecliptic occur within $\sim10^\circ$ in both longitude and latitude of the parent AR. The remaining three cases comprise: one AR with no open lines; one large filament eruption; and one event in which the AR is absent from the synoptic map. While PFSS model is a simple approximation of the global coronal field, these results suggest that, in most events, an open-field path likely exists for seed ${}^3\mathrm{He}$ to the same-event shock.

\subsection{Solar sources}  \label{subsec:sources}

As shown in Table~\ref{tab:source}, 12 of the 23 events were associated with jet-like ejections in the parent AR. These jets were associated with the most intense flares in our sample, including X-class X-ray flares, while events lacking jets were limited to C- and M-class flares and exhibited no X-class activity.

\begin{figure}
\centering
\begin{interactive}{animation}{fig_8_event_14.mp4}
    \includegraphics[width=0.9\textwidth]{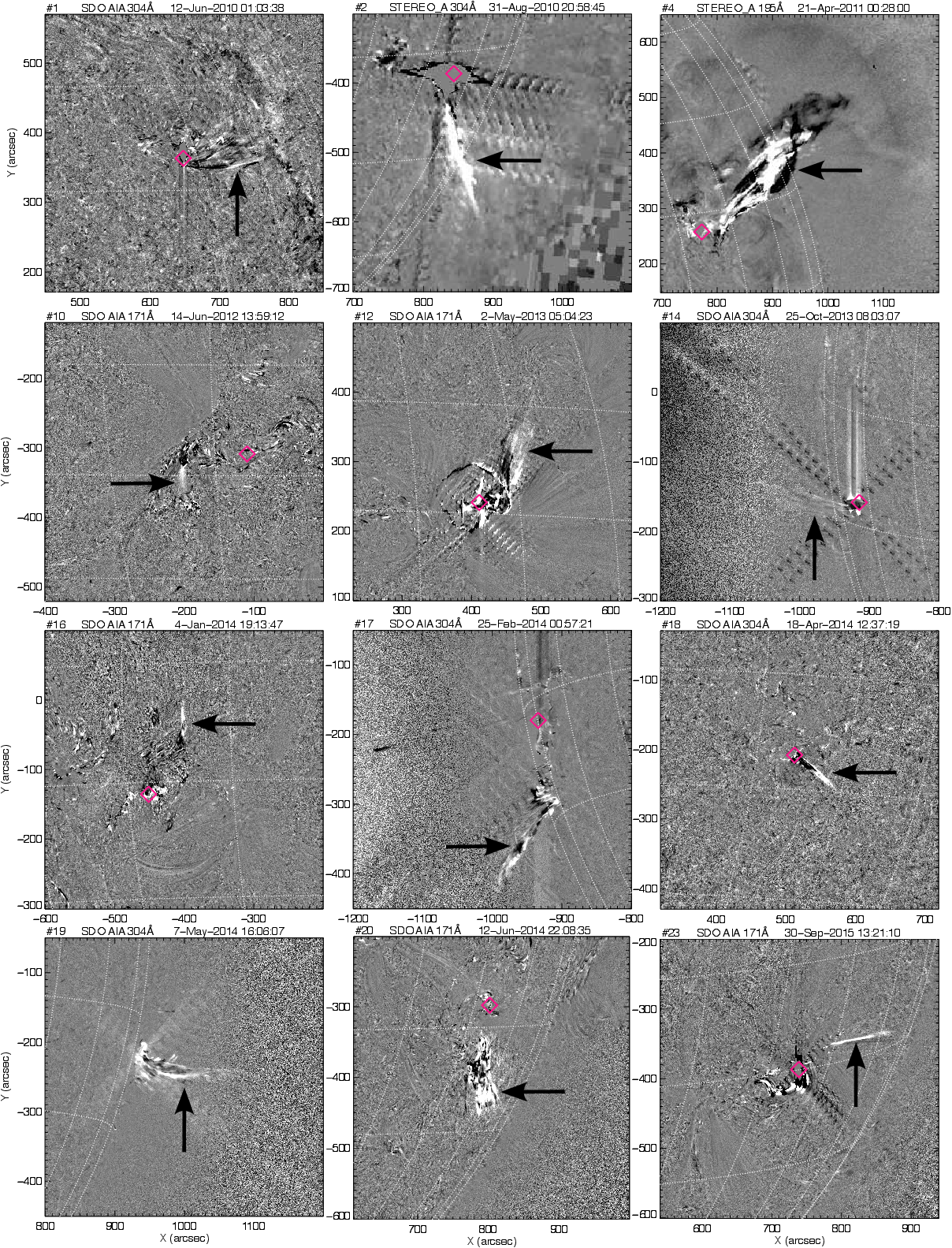}
  \end{interactive}
\caption{EUV images of source flares in jet-associated events. Panels show 1-min running-difference SDO/AIA 304\,{\AA}  or 171\,{\AA} (events \#1, 10, 12, 14, 16--20, 23) and 2.5-min STEREO-A/EUVI 304\,{\AA} or 195\,{\AA} (events \#2, 4). Jets are marked by arrows; a 10$^\circ$ Stonyhurst grid is overlaid. Flare locations are magenta diamonds; in \#19 the flare is behind the west limb as seen from SDO. An animation for panel~\#14 accompanies the online article. The animation starts on October 25th, 2013 at 07:56:07 and ends the same day at 08:09:55. The real-time duration of this animation is 3 seconds. Animations for all panels are available in Zenodo at \dataset[doi: 10.5281/zenodo.17299168]{https://doi.org/10.5281/zenodo.17299168}.
\label{fig:jets}}
\end{figure}

Figure~\ref{fig:jets} presents the source flares for the 12 events accompanied by EUV jets. Snapshots are shown at times when the jets were clearly visible, $\sim10$ minutes after the onset of the associated type III radio bursts. The jets were identified via visual inspection of time sequences of EUV images from SDO and STEREO at multiple wavelengths. For SDO/AIA observations, jets were most clearly visible in 171\,{\AA} or 304\,{\AA} channels. For two far-side events, the highest-cadence images from STEREO/EUVI at 195\,{\AA} or 304\,{\AA} are shown. Vertical bright stripes and diagonal patterns in some panels result from detector saturation and diffraction fringes caused by strong EUV emission. These jets were often accompanied by additional activity in the AR, such as large-scale coronal propagating fronts, broad eruptions, or localized brightening (see the animation associated with Fig.~\ref{fig:jets} in the online journal; additional per-panel animations are archived at Zenodo, see caption). The morphology of the jets varied considerably across events, ranging from simple, straight jets (\#2, 10, 16, 18, 23), to fan-shaped jets (\#1, 12, 14, 20), and helical jets (\#4, 17, 19). In events \#10, 16, 17 and 23, the jets originated at the footpoints of loops, spatially distinct from the main flare brightening. 

In event \#4, a dome-shaped front (a darker arc) enveloping the helical jet has been interpreted as a shock wave \citep[e.g.,][]{2010ApJ...716L..57V}. Notably, this event exhibited the highest ${}^3\mathrm{He}/{}^4\mathrm{He}$ ratio in the survey, approximately 40 times the survey average. Helical jets observed in the sources of ISEP events have been previously associated with exceptional strong enhancement of ${}^3\mathrm{He}$ and heavy ions \citep{2015ApJ...806..235N,2018ApJ...852...76B,2018ApJ...869L..21B}. 

Appendix~\ref{sec:nojets} shows for comparison activity in the solar source of events where no jets were observed.

\subsection{Residual ${}^3\mathrm{He}$}  \label{subsec:res}

\begin{figure}
\centering
\epsscale{1.1}
\plotone{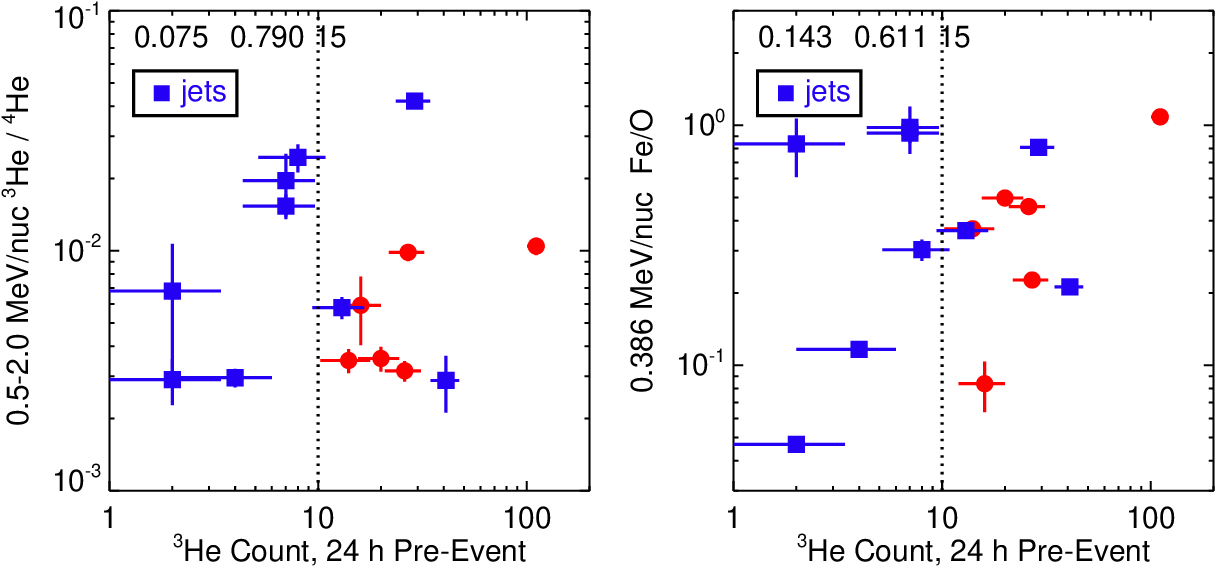}
\caption{Ratios of ${}^3\mathrm{He}/{}^4\mathrm{He}$ and Fe/O as a function of the total 0.3--0.5\,MeV\,nucleon$^{-1}$ ${}^3\mathrm{He}$ count integrated over the 24 hrs preceding the onset of the ${}^3\mathrm{He}$ intensity increase. Blue squares indicate events associated with jets at the source; red circles denote events where no jets were observed.
\label{fig:res}}
\end{figure}

Sporadic ${}^3\mathrm{He}$ counts observed prior to proton events indicate the presence of residual material available for further acceleration. Figure~\ref{fig:res} shows the ${}^3\mathrm{He}/{}^4\mathrm{He}$ and Fe/O as a function of the total suprathermal 0.3--0.5\,MeV\,nucleon$^{-1}$ integrated ${}^3\mathrm{He}$ count over the 24 hours preceding the onset of the ${}^3\mathrm{He}$ increase associated with the proton event. No correlation is found between the elemental ratios and the pre-event integrated ${}^3\mathrm{He}$ count, implying that the residual population does not by itself set the observed ${}^3\mathrm{He}$ enhancement. Interestingly, events with lower pre-event ${}^3\mathrm{He}$ counts ($\lesssim10$) tend to be associated with jets. The low pre-event background suggests little remnant material, yet these GSEP events are ${}^3\mathrm{He}$-enriched, consistent with the jet supplying the seed ${}^3\mathrm{He}$. In contrast, events lacking visible jets at the source show higher pre-event ${}^3\mathrm{He}$ counts ($\gtrsim10$), pointing to a remnant suprathermal population as a likely contributor. The lowest pre-event ${}^3\mathrm{He}$ counts occur in events \#2 and \#10, while the highest occur in event \#6.   

\subsection{Association with CME speed}  \label{subsec:cme}

\begin{figure}
\centering
\epsscale{1.1}
\plotone{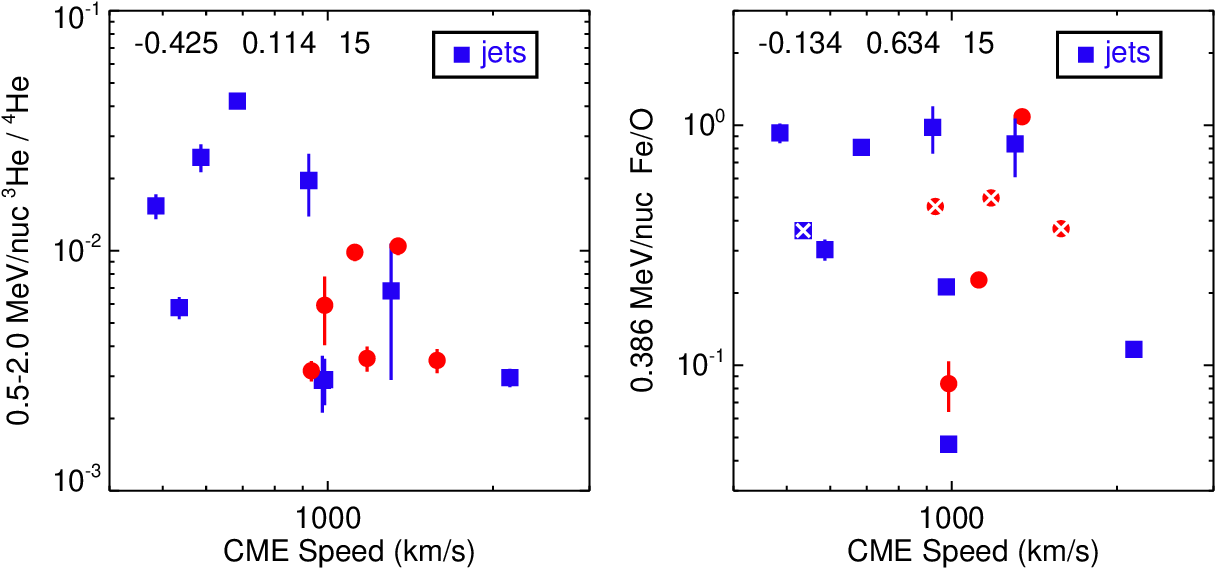}
\caption{${}^3\mathrm{He}/{}^4\mathrm{He}$ and Fe/O ratios versus CME speed. Blue squares indicate events associated with jets at the source, while red circles denote events where no jets were observed. Points marked with white crosses (right panel), as well as the the three numerical values shown, have the same meaning as in Fig.~\ref{fig:hist-abu}.
\label{fig:cme}}
\end{figure}

A moderate, negative correlation between ${}^3\mathrm{He}/{}^4\mathrm{He}$ ratio and CME (plane-of-sky) speed is apparent in Fig.~\ref{fig:cme}(left). By contrast, Fe/O shows no significant trend with speed (Fig.~\ref{fig:cme}, right), consistent with \citet[][$r\approx -0.11$]{2006ApJ...649..470D}. The absence of an Fe/O trend is expected if Fe/O is strongly affected by $Q/M$-dependent shock acceleration \citep[e.g.,][]{2016ApJ...816...68D}, which can mask any simple speed dependence. Events with jets span a broader range of CME speeds than those without jets, possibly reflecting different partitions of energy between the jet and the CME at the source.

\section{Discussion}  \label{sec:disc}

\citet{2006ApJ...649..470D} reported a correlation of $r = 0.66$ between ${}^3\mathrm{He}/{}^4\mathrm{He}$ and Fe/O based on 29 GSEP events with finite ${}^3\mathrm{He}$ peaks. However, excluding two outliers reduced the correlation to $r = 0.026$, leading the authors to concluded that Fe/O is independent of ${}^3\mathrm{He}/{}^4\mathrm{He}$ in GSEP events. Similar findings were reported for ISEP events by \citet{1986ApJ...303..849M} and \citet{1994ApJS...90..649R}. \citet{2022ApJS..263...22H} show that ${}^3\mathrm{He}$-rich time periods with low Fe/O, indicative of mixed ISEP and GSEP events, also tend to exhibit lower ${}^3\mathrm{He}/{}^4\mathrm{He}$, whereas ${}^3\mathrm{He}/{}^4\mathrm{He}$ and Fe/O in pure ISEP events are uncorrelated. The present observations show that ${}^3\mathrm{He}/{}^4\mathrm{He}$ and Fe/O are positively correlated in GSEP events as suggested in \citet{2022ApJS..263...22H}. The correlation is strongest for GSEP events associated with concurrent jets, consistent with both ratios increasing as the fraction of jet-supplied seed ions (${}^3\mathrm{He}$, ${}^4\mathrm{He}$, O, Fe) that are subsequently re-accelerated becomes larger. Compared with the broader GSEP set of \citet{2006ApJ...649..470D}, which exhibited up to two orders of magnitude event-to-event variation in elemental ratios, our focus on ${}^3\mathrm{He}$-enriched GSEP events reduces scatter and may help reveal this correlation.

If suprathermal ions are supplied by the parent flare, well-connected events should exhibit higher enrichments. Consistent with this expectation, we find a negative correlation between Fe/O and the absolute longitudinal separation between the magnetic footpoint and the associated flare. In contrast, ${}^3\mathrm{He}/{}^4\mathrm{He}$ shows no correlation. This trend persists when the most distant event (\#14) is omitted, reinforcing that the contrasting behaviors of ${}^3\mathrm{He}/{}^4\mathrm{He}$ (no clear dependence) and Fe/O (moderate anti-correlation) are not driven by a single point. The ${}^3\mathrm{He}/{}^4\mathrm{He}$ ratio in ISEP events spans over three orders of magnitude, whereas Fe/O varies by only $\sim1.5$ orders \citep[e.g.,][]{2022ApJS..263...22H}. This suggests that Fe/O may retain some dependence on longitudinal separation, whereas the large event-to-event dispersion in ${}^3\mathrm{He}/{}^4\mathrm{He}$ likely masks any such trend. 

Identified jets coincided with type III radio bursts, a characteristic feature of ISEP events \citep{1986ApJ...308..902R,2006ApJ...650..438N}. Jets, as signatures of magnetic reconnection involving field lines open to IP space \citep[e.g.,][]{1992PASJ...44L.173S}, have been firmly associated with the origin of ${}^3\mathrm{He}$-rich SEPs \citep[e.g.,][]{2001ApJ...562..558K,2006ApJ...639..495W,2006ApJ...648.1247P,2006ApJ...650..438N,2008ApJ...675L.125N,2015ApJ...806..235N,2023FrASS..1048467N,2015AA...580A..16C,2016AN....337.1024I,2016ApJ...823..138M,2020SSRv..216...24B,2021AA...656L..11B,2023AA...673L...5B,2025ApJ...981..178B,2020ApJS..246...42W,2024ApJ...975...84L}. Jet ejections may supply the ${}^3\mathrm{He}$-rich seed population, which can then be further accelerated by CME-driven shocks during the same proton event (\citealp{2005GeoRL..32.2101L}; \citealp{2020SSRv..216...20R}; \citealp{2024ApJ...974..220H}).

Of the 15 events in our final sample (see Section~\ref{subsec:element}), nine (events \#1, 2, 10, 14, 16, 17, 19, 20, 23) showed EUV jets in the parent AR, alongside other AR activity. Notably, the four events with the highest ${}^3\mathrm{He}/{}^4\mathrm{He}$ ratios (\#1, 14, 19, 20) all belong to this jet-associated subset (see Fig.~\ref{fig:hist-abu}, right panel). This complete overlap highlights jets as a likely co-temporal source of the ${}^3\mathrm{He}$-rich seed population in GSEP events. The elevated ${}^3\mathrm{He}/{}^4\mathrm{He}$ ratio in event \#14 may reflect contributions from two EUV jets (Table~\ref{tab:source}) within the integration window. Compared to ISEP events, those examined here were linked to more energetic flares, so some jets may have been missed due to observational limitations. 

No clear correlation was found between the amount of residual suprathermal ${}^3\mathrm{He}$ and the observed ${}^3\mathrm{He}/{}^4\mathrm{He}$ or Fe/O ratios. Similar results were reported by \citet{2006ApJ...649..470D}, who noted a lack of correlation between 8-day pre-event suprathermal Fe/O and the Fe/O ratio during GSEP events. In this study, we use a 24-hr integration window for suprathermal ${}^3\mathrm{He}$ to remain consistent with our event selection. Longer windows can fold in ions from prior GSEP events. Indeed, \citet{2006ApJ...649..470D} excluded ~30\% of events because their 8-day interval overlapped previous GSEP events. Although the ratios do not track the pre-event counts, the distribution of counts by jet classification (Fig.~\ref{fig:res}) suggests two regimes: jet events generally have low pre-event ${}^3\mathrm{He}$, whereas non-jet events often have higher pre-event ${}^3\mathrm{He}$, implying different relative contributions of contemporaneous jet seeds and residual populations.

Finally, we note a moderate anti-correlation between ${}^3\mathrm{He}/{}^4\mathrm{He}$ ratio and CME speed, consistent with a scenario in which the jet or residual pool supplies the ${}^3\mathrm{He}$-rich seed population, while the CME-driven shock increasingly accelerates ambient ${}^4\mathrm{He}$ as the speed rises, thereby reducing the final ${}^3\mathrm{He}/{}^4\mathrm{He}$. Because the injection threshold depends on both shock speed and shock obliquity \citep[e.g.,][]{2013A&A...558A.110B}, a mixture of geometries in the sample will naturally weaken any global ${}^3\mathrm{He}/{}^4\mathrm{He}$ speed trend.

\section{Conclusion}  \label{sec:con}

We investigated the origin of ${}^3\mathrm{He}$ enrichment in 23 high-energy (25--50\,MeV) proton events. After excluding intervals contaminated by independent ISEP events, and removing two events with ISEP characteristics, we retained 15 clean GSEP events in which ${}^3\mathrm{He}$ abundance enhancement is consistent with re-acceleration of suprathermal flare ions. Our main findings for these 15 events are:

\begin{itemize}
 \item{EUV imaging shows jets in $\sim60$\% (9/15) of the GSEP sources. The four largest ${}^3\mathrm{He}/{}^4\mathrm{He}$ ratios all occur in jet-associated events, consistent with jets supplying the ${}^3\mathrm{He}$-rich seed population.}
 \item{${}^3\mathrm{He}/{}^4\mathrm{He}$ and Fe/O are positively correlated in the GSEP events, with the correlation driven by jet-associated sources. This supports a scenario in which jets provide a common seed population that is subsequently re-accelerated by the CME shock.}
 \item{GSEP events with jets generally show lower pre-event ${}^3\mathrm{He}$ counts, whereas events without jets show higher counts. This pattern suggests that jets can supply ${}^3\mathrm{He}$ locally at the time of the event, while in non-jet cases a residual suprathermal population from prior activity may contribute. Neither source alone controls the elemental ratios.}
 \item{Well-connected GSEP events tend to show higher Fe/O ratios, whereas a similar trend is not evident in ${}^3\mathrm{He}/{}^4\mathrm{He}$. Plausibly the large intrinsic variability of the suprathermal helium seed population masks any magnetic connection dependence.}
 \item{Heavy-ion abundances in the ${}^3\mathrm{He}$-enriched GSEP events are consistent with average GSEP values at comparable energies, suggesting that the heavy-ion component is largely drawn from the ambient suprathermal seed population and then re-accelerated by the CME shock.}
\end{itemize}

About 40\% (9/21) of the apparent ${}^3\mathrm{He}$ enrichments were due to overlap with independent ISEP activity, five whole events and brief intervals in four others, showing that a sizeable fraction of ${}^3\mathrm{He}$-rich GSEP identifications can be superpositions rather than re-acceleration. This study provides a framework for further investigation: extending the live catalog beyond 2021 March 1 and incorporating measurements at different heliocentric distances (Parker Solar Probe, Solar Orbiter) will better constrain when and how ${}^3\mathrm{He}$ enrichment arises in GSEP events.

\begin{acknowledgments}
R.B. acknowledges support by NASA grants 80NSSC21K1316 and 80NSSC22K0757. ACE work at JHU/APL is supported by NASA contract 80NSSC22K0374. The LASCO C2 CME catalog is generated and maintained at the CDAW Data Center by NASA and The Catholic University of America in cooperation with the Naval Research Laboratory. SOHO is a project of international cooperation between ESA and NASA.
\end{acknowledgments}

\software{SolarSoft \citep{1998SoPh..182..497F}
          }

\appendix

\section{CME in event \#23}  \label{sec:cme23}

\begin{figure}
\centering
\epsscale{1.1}
\plotone{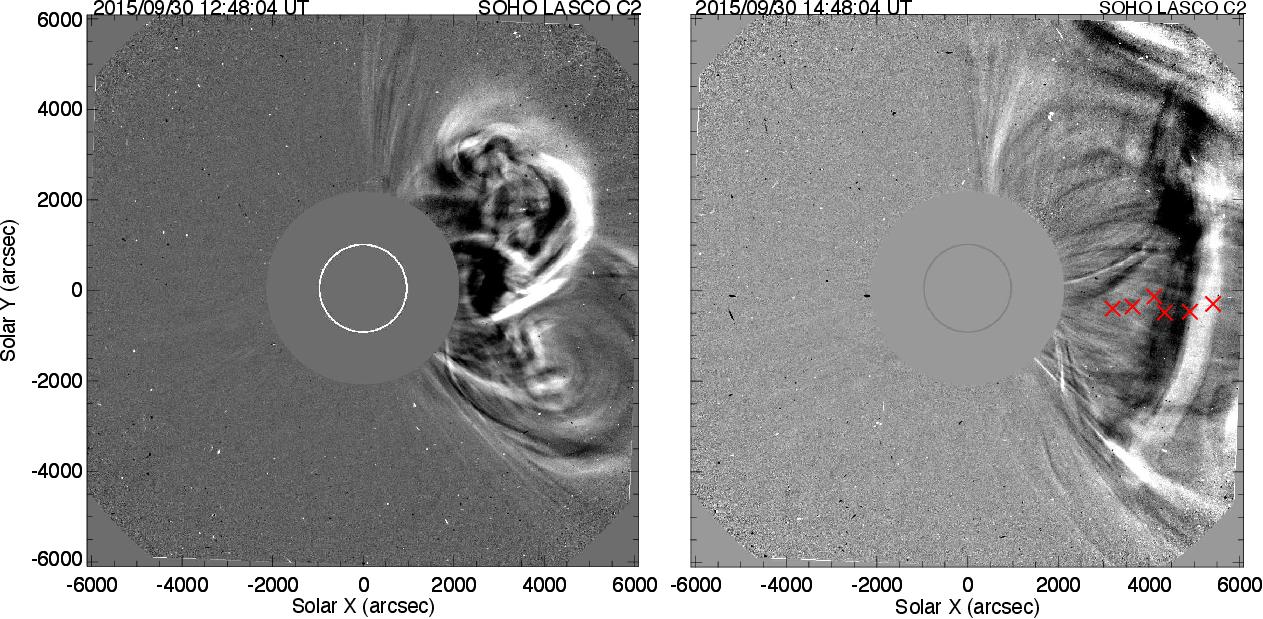}
\caption{LASCO C2 running difference images. Left: The two adjacent CMEs, shown 30 minutes before X-ray flare associated with event \#23. Right: Tracing of the CME front associated with event \#23. Each red cross corresponds to the average from several points manually selected along the outermost bright front at a given time. 
\label{fig:cme_both}}
\end{figure}

\begin{figure}
\centering
\epsscale{.55}
\plotone{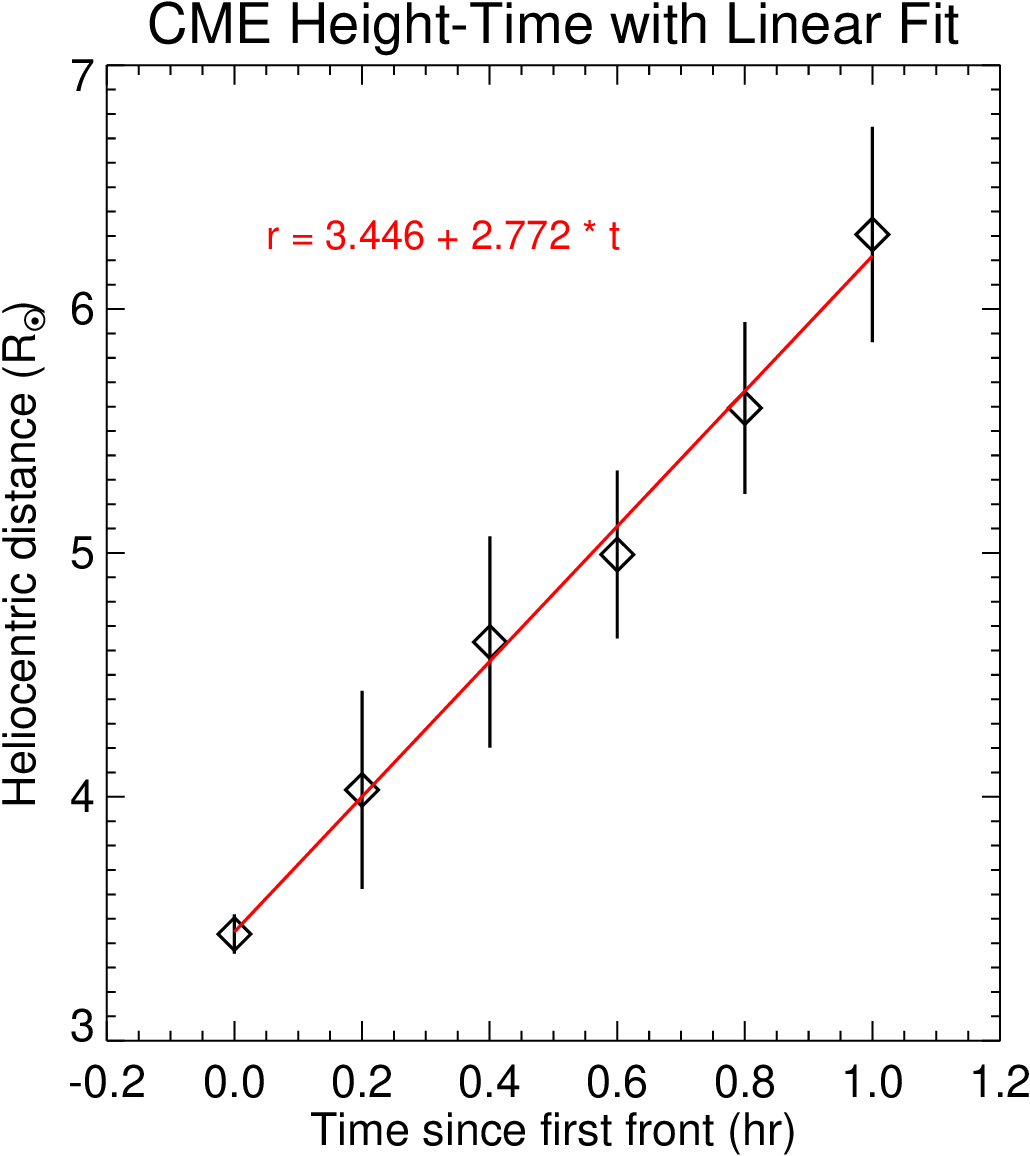}
\caption{Mean heliocentric distance ($r$) of the CME front vs. time ($t$) since the first front. Red line presents the linear fit; the slope is directly proportional to the CME speed.
\label{fig:front}}
\end{figure}

The LASCO CME catalog reports two consecutive CMEs on 2015 September 30, with the first appearances in the C2 coronagraph at 09{:}36\,UT and 10{:}00\,UT, respectively. Both were wide (102$^\circ$ and 128$^\circ$) with speeds of 586 and 602\,km\,s$^{-1}$. The CMEs erupted westward, next to each other, with position angles separated by only $\sim50^\circ$. Event \#23 was associated with an M1.1 X-ray flare at 13{:}18\,UT. The two preceding CMEs can be seen clearly by LASCO C2 in Figure~\ref{fig:cme_both}(Left) taken at 12{:}48\,UT, 30 minutes before M1.1 X-ray flare start time. At 13{:}48\,UT, a new large-scale structure can be seen emerging in the wake of the two preceding CMEs. We track the bright front from this CME until it exits the field-of-view of LASCO C2 (Figure~\ref{fig:cme_both}, Right). We measure its plane-of-sky speed to be approximately 536\,km\,s$^{-1}$ fitting the CME height-time plot (Figure~\ref{fig:front}). Note, there are no large-scale CMEs emerging after 13{:}48\,UT. 

\section{Solar sources without jets} \label{sec:nojets}

\begin{figure}
\centering
\begin{interactive}{animation}{fig_13_event_6.mp4}
    \includegraphics[width=0.9\textwidth]{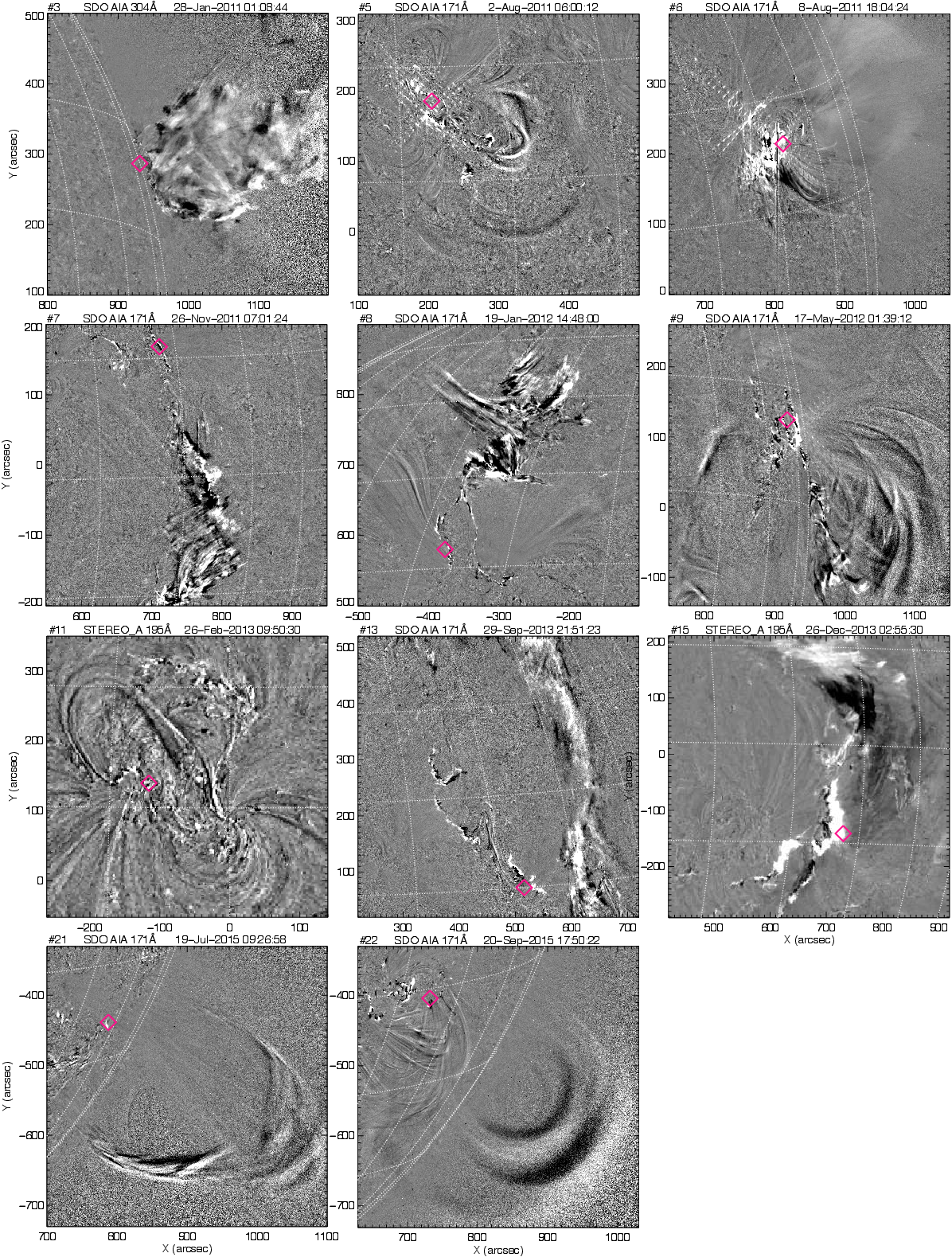}
  \end{interactive}
\caption{EUV images of source flares in events where no jets were observed. Panels show 1-min running-difference SDO/AIA 304\,{\AA}  or 171\,{\AA} (events \#3, 5--9, 13, 21--22) and 5-min STEREO-A/EUVI 195\,{\AA} (events \#11, 15). A 10$^\circ$ Stonyhurst grid is overlaid. Flare locations are magenta diamonds. An animation for panel~\#6 accompanies the online article. The animation starts on August 8th, 2011 at 18:00:00 and ends the same day at 08:13:50. The real-time duration of this animation is 3 seconds. Animations for all panels are available in Zenodo at \dataset[doi: 10.5281/zenodo.17299730]{https://doi.org/10.5281/zenodo.17299730}.
\label{fig:nojets}}
\end{figure}

Figure~\ref{fig:nojets} displays activity in the solar source of events where jets were not observed.

\bibliography{ads}{}
\bibliographystyle{aasjournalv7}

\end{document}